\documentclass[longauth]{aa}  

\usepackage{graphicx}
\usepackage{caption, subcaption}
\usepackage{xcolor}
\usepackage{txfonts}

\usepackage[]{hyperref}
\usepackage[normalem]{ulem}
\usepackage{orcidlink}
\usepackage{booktabs}

\newcounter{pta}

\newcommand{\epta}[1]{{EPTA~D.R.~1}}
\newcommand{\ipta}[1]{{IPTA~D.R.~1}}

\def\wm{4}

\begin{document}

   \title{The second data release from the European Pulsar Timing Array}
   \subtitle{V. Search for continuous gravitational wave signals}

\author{{EPTA Collaboration and InPTA Collaboration:}
    J.~Antoniadis\orcidlink{0000-0003-4453-776}\inst{\ref{forth}, \ref{mpifr}}, 
    \ifnum\wm>1 P.~Arumugam\orcidlink{0000-0001-9264-8024}\inst{\ref{IITR}}, \fi
    \ifnum\wm>1 S.~Arumugam\orcidlink{0009-0001-3587-6622}\inst{\ref{IITH_El}}, \fi
    \ifnum\wm=5 P.~Auclair\orcidlink{0000-0002-4814-1406}\inst{\ref{curl}}, \fi
    S.~Babak\orcidlink{0000-0001-7469-4250}\inst{\ref{apc}}\thanks{stas@apc.in2p3.fr}, 
    \ifnum\wm>1 M.~Bagchi\orcidlink{0000-0001-8640-8186}\inst{\ref{IMSc}, \ref{HBNI}}, \fi
    A.-S.~Bak~Nielsen\orcidlink{ 0000-0002-1298-9392}\inst{\ref{mpifr}, \ref{unibi}}, 
    \ifnum\wm=5 E.~Barausse\orcidlink{0000-0001-6499-6263 }\inst{\ref{sissa}}, \fi
    C.~G.~Bassa\orcidlink{0000-0002-1429-9010}\inst{\ref{astron}}, 
    \ifnum\wm>1 A.~Bathula\orcidlink{0000-0001-7947-6703} \inst{\ref{IISERM}}, \fi
    A.~Berthereau\inst{\ref{lpc2e}, \ref{nancay}}, 
    M.~Bonetti\orcidlink{0000-0001-7889-6810}\inst{\ref{unimib}, \ref{infn-unimib}, \ref{inaf-brera}}, 
    E.~Bortolas\inst{\ref{unimib}, \ref{infn-unimib}, \ref{inaf-brera}}, 
    P.~R.~Brook\orcidlink{0000-0003-3053-6538}\inst{\ref{unibir}}, 
    M.~Burgay\orcidlink{0000-0002-8265-4344}\inst{\ref{inaf-oac}}, 
    R.~N.~Caballero\orcidlink{0000-0001-9084-9427}\inst{\ref{HOU}}, 
    \ifnum\wm=5 C.~Caprini\orcidlink{0000-0001-5393-2205}\inst{\ref{unige}, \ref{CERN}}, \fi
    A.~Chalumeau\orcidlink{0000-0003-2111-1001}\inst{\ref{unimib}}\ifnum\wm=2\thanks{aurelien.chalumeau@unimib.it}\fi, 
    D.~J.~Champion\orcidlink{0000-0003-1361-7723}\inst{\ref{mpifr}}, 
    S.~Chanlaridis\orcidlink{0000-0002-9323-9728}\inst{\ref{forth}}, 
    S.~Chen\orcidlink{0000-0002-3118-5963}\inst{\ref{kiaa}}\ifnum\wm=3\thanks{sychen@pku.edu.cn}\fi, 
    I.~Cognard\orcidlink{0000-0002-1775-9692}\inst{\ref{lpc2e}, \ref{nancay}}, 
    \ifnum\wm=5 M.~Crisostomi\orcidlink{0000-0002-7622-4911}\inst{\ref{sissa}}, \fi
    \ifnum\wm>1 S.~Dandapat\orcidlink{0000-0003-4965-9220}\inst{\ref{TIFR}}, \fi
    \ifnum\wm>1 D.~Deb\orcidlink{0000-0003-4067-5283}\inst{\ref{IMSc}},  \fi
    \ifnum\wm>1 S.~Desai\orcidlink{0000-0002-0466-3288}\inst{\ref{IITH_Ph}}, \fi
    G.~Desvignes\orcidlink{0000-0003-3922-4055}\inst{\ref{mpifr}}, 
    \ifnum\wm>1 N.~Dhanda-Batra \inst{\ref{UoD}}, \fi
    \ifnum\wm>1 C.~Dwivedi\orcidlink{0000-0002-8804-650X}\inst{\ref{IIST}}, \fi
    M.~Falxa\inst{\ref{apc}, \ref{lpc2e}}\ifnum\wm=4\thanks{falxa@apc.in2p3.fr}\fi, 
    \ifnum\wm=5 F.~Fastidio\inst{\ref{surrey}, \ref{unimib},}, \fi
    \ifnum\wm=4 I.~Ferranti\orcidlink{0009-0000-1575-2051}\inst{\ref{unimib}, \ref{apc}}, \fi 
    R.~D.~Ferdman\inst{\ref{uea}}, 
    A.~Franchini\orcidlink{0000-0002-8400-0969}\inst{\ref{unimib}, \ref{infn-unimib}}, 
    J.~R.~Gair\orcidlink{0000-0002-1671-3668}\inst{\ref{aei}}, 
    B.~Goncharov\orcidlink{0000-0003-3189-5807}\inst{\ref{gssi}, \ref{lngs}}
    \ifnum\wm>1 A.~Gopakumar\orcidlink{0000-0003-4274-4369}\inst{\ref{TIFR}}, \fi
    E.~Graikou\inst{\ref{mpifr}}, 
    J.-M.~Grie{\ss}meier\orcidlink{0000-0003-3362-7996}\inst{\ref{lpc2e}, \ref{nancay}}, 
    \ifnum\wm=5 A.~Gualandris\orcidlink{0000-0002-9420-2679}\inst{\ref{surrey}}, \fi
    L.~Guillemot\orcidlink{0000-0002-9049-8716}\inst{\ref{lpc2e}, \ref{nancay}}, 
    Y.~J.~Guo\inst{\ref{mpifr}}\ifnum\wm=3\thanks{yjguo@mpifr-bonn.mpg.de}\fi, 
    \ifnum\wm>1 Y.~Gupta\orcidlink{0000-0001-5765-0619}\inst{\ref{NCRA}}, \fi
    \ifnum\wm>1 S.~Hisano\orcidlink{0000-0002-7700-3379}\inst{\ref{KU_J}}, \fi
    H.~Hu\orcidlink{0000-0002-3407-8071}\inst{\ref{mpifr}},  
    F.~Iraci\inst{\ref{unica}\ref{inaf-oac}}, 
    D.~Izquierdo-Villalba\orcidlink{0000-0002-6143-1491}\inst{\ref{unimib}, \ref{infn-unimib}}, 
    J.~Jang\orcidlink{0000-0003-4454-0204}\inst{\ref{mpifr}}\ifnum\wm=1\thanks{jjang@mpifr-bonn.mpg.de}\fi, 
    J.~Jawor\orcidlink{0000-0003-3391-0011}\inst{\ref{mpifr}}, 
    G.~H.~Janssen\orcidlink{0000-0003-3068-3677}\inst{\ref{astron}, \ref{imapp}}, 
    A.~Jessner\orcidlink{0000-0001-6152-9504}\inst{\ref{mpifr}}, 
    \ifnum\wm>1 B.~C.~Joshi\orcidlink{0000-0002-0863-7781}\inst{\ref{NCRA}, \ref{IITR}}, \fi
    \ifnum\wm>1 F.~Kareem\orcidlink{0000-0003-2444-838X} \inst{\ref{IISERK}, \ref{CESSI}}, \fi
    R.~Karuppusamy\orcidlink{0000-0002-5307-2919}\inst{\ref{mpifr}}, 
    E.~F.~Keane\orcidlink{0000-0002-4553-655X}\inst{\ref{tcd}}, 
    M.~J.~Keith\orcidlink{0000-0001-5567-5492}\inst{\ref{jbca}}\ifnum\wm=2\thanks{michael.keith@manchester.ac.uk}\fi, 
    \ifnum\wm>1 D.~Kharbanda\orcidlink{0000-0001-8863-4152}\inst{\ref{IITH_Ph}}, \fi
    \ifnum\wm=5 T.~Khizriev, \inst{\ref{SAI}}, \fi
    \ifnum\wm>1 T.~Kikunaga\orcidlink{0000-0002-5016-3567} \inst{\ref{KU_J}}, \fi
    \ifnum\wm>1 N.~Kolhe\orcidlink{0000-0003-3528-9863} \inst{\ref{XCM}}, \fi
    M.~Kramer\inst{\ref{mpifr}, \ref{jbca}}, 
    M.~A.~Krishnakumar\orcidlink{0000-0003-4528-2745}\inst{\ref{mpifr}, \ref{unibi}}, 
    K.~Lackeos\orcidlink{0000-0002-6554-3722}\inst{\ref{mpifr}}, 
    K.~J.~Lee\inst{\ref{pku}, \ref{naoc}}, 
    K.~Liu\inst{\ref{mpifr}}\ifnum\wm=1\thanks{kliu@mpifr-bonn.mpg.de}\fi, 
    Y.~Liu\orcidlink{0000-0001-9986-9360}\inst{\ref{naoc},  \ref{unibi}}, 
    A.~G.~Lyne\inst{\ref{jbca}}, 
    J.~W.~McKee\orcidlink{0000-0002-2885-8485}\inst{\ref{milne}, \ref{daim}}, 
    \ifnum\wm>1 Y.~Maan\inst{\ref{NCRA}}, \fi
    R.~A.~Main\inst{\ref{mpifr}}, 
    \ifnum\wm=4 S.~Manzini\orcidlink{0009-0005-1149-5330}\inst{\ref{unimib}, \ref{apc}}, \fi
    M.~B.~Mickaliger\orcidlink{0000-0001-6798-5682}\inst{\ref{jbca}}, 
    \ifnum\wm=5 H.~Middleton\orcidlink{0000-0001-5532-3622}\inst{\ref{unibir}}, \fi
    \ifnum\wm=5 A.~Neronov\inst{\ref{apc}, \ref{EPFL}}, \fi
    I.~C.~Ni\c{t}u\orcidlink{0000-0003-3611-3464}\inst{\ref{jbca}}, 
    \ifnum\wm>1 K.~Nobleson\orcidlink{0000-0003-2715-4504}\inst{\ref{BITS}}, \fi
    \ifnum\wm>1 A.~K.~Paladi\orcidlink{0000-0002-8651-9510}\inst{\ref{IISc}}, \fi
    A.~Parthasarathy\orcidlink{0000-0002-4140-5616}\inst{\ref{mpifr}}\ifnum\wm=2\thanks{aparthas@mpifr-bonn.mpg.de}\fi, 
    B.~B.~P.~Perera\orcidlink{0000-0002-8509-5947}\inst{\ref{arecibo}}, 
    D.~Perrodin\orcidlink{0000-0002-1806-2483}\inst{\ref{inaf-oac}}, 
    A.~Petiteau\orcidlink{0000-0002-7371-9695}\inst{\ref{irfu}, \ref{apc}ee}, 
    N.~K.~Porayko\inst{\ref{unimib}, \ref{mpifr}}\ifnum\wm=5\thanks{nataliya.porayko@unimib.it}\fi, 
    A.~Possenti\inst{\ref{inaf-oac}}, 
    \ifnum\wm>1 T.~Prabu\inst{\ref{RRI}}, \fi
    \ifnum\wm=5 K.~Postnov\orcidlink{0000-0002-1705-617X}\inst{\ref{SAI},  \ref{KFU}}, \fi
    H.~Quelquejay~Leclere\inst{\ref{apc}}\ifnum\wm=5\thanks{quelquejay@apc.in2p3.fr}\fi,
    \ifnum\wm>1 P.~Rana\orcidlink{0000-0001-6184-5195}\inst{\ref{TIFR}}, \fi
    \ifnum\wm=5 A.~Roper Pol\orcidlink{0000-0003-4979-4430}\inst{\ref{unige}}, \fi
    A.~Samajdar\orcidlink{0000-0002-0857-6018}\inst{\ref{uni-potsdam}}, 
    S.~A.~Sanidas\inst{\ref{jbca}}, 
    \ifnum\wm=5 D.~Semikoz\inst{\ref{apc}}, \fi
    A.~Sesana\inst{\ref{unimib}, \ref{infn-unimib}, \ref{inaf-brera}}\ifnum\wm=5\thanks{alberto.sesana@unimib.it}\fi, 
    G.~Shaifullah\orcidlink{0000-0002-8452-4834}\inst{\ref{unimib}, \ref{infn-unimib}, \ref{inaf-oac}}\ifnum\wm=1\thanks{golam.shaifullah@unimib.it}\fi, 
    \ifnum\wm>1 J.~Singha\orcidlink{0000-0002-1636-9414}\inst{\ref{IITR}}, \fi
    \ifnum\wm=5 C.~Smarra\inst{\ref{sissa}}, \fi
    L.~Speri\orcidlink{0000-0002-5442-7267}\inst{\ref{aei}}\ifnum\wm=4\thanks{lorenzo.speri@aei.mpg.de}\fi, 
    R.~Spiewak\inst{\ref{jbca}}, 
    \ifnum\wm>1 A.~Srivastava\orcidlink{0000-0003-3531-7887} \inst{\ref{IITH_Ph}}, \fi
    B.~W.~Stappers\inst{\ref{jbca}}, 
    \ifnum\wm=5 D.~A.~Steer\orcidlink{0000-0002-8781-1273}\inst{\ref{apc}}\fi
    \ifnum\wm>1 M.~Surnis\orcidlink{0000-0002-9507-6985}\inst{\ref{IISERB}}, \fi
    S.~C.~Susarla\orcidlink{0000-0003-4332-8201}\inst{\ref{uog}}, 
    \ifnum\wm>1 A.~Susobhanan\orcidlink{0000-0002-2820-0931}\inst{\ref{CGCA}}, \fi
    \ifnum\wm>1 K.~Takahashi\orcidlink{0000-0002-3034-5769}\inst{\ref{KU_J1}, \ref{KU_J2}} \fi
    \ifnum\wm>1 P.~Tarafdar\orcidlink{0000-0001-6921-4195}\inst{\ref{IMSc}}\fi
    G.~Theureau\orcidlink{0000-0002-3649-276X}\inst{\ref{lpc2e},  \ref{nancay},  \ref{luth}}, 
    C.~Tiburzi\inst{\ref{inaf-oac}}, 
    \ifnum\wm=5 R.~J.~Truant\inst{\ref{unimib}}, \fi
    E.~van~der~Wateren\orcidlink{0000-0003-0382-8463}\inst{\ref{astron}, \ref{imapp}}, 
    \ifnum\wm=5 S.~Valtolina\inst{\ref{aei2}}, \fi
    A.~Vecchio\orcidlink{0000-0002-6254-1617}\inst{\ref{unibir}}, 
    V.~Venkatraman~Krishnan\orcidlink{0000-0001-9518-9819}\inst{\ref{mpifr}}, 
    J.~P.~W.~Verbiest\orcidlink{0000-0002-4088-896X}\inst{\ref{FSI}, \ref{unibi}, \ref{mpifr}}, 
    J.~Wang\orcidlink{0000-0003-1933-6498}\inst{\ref{unibi},  \ref{airub},  \ref{bnuz}}, 
    L.~Wang\inst{\ref{jbca}} and 
    Z.~Wu\orcidlink{0000-0002-1381-7859}\inst{\ref{naoc}, \ref{unibi}}.
    }

\institute{
{Institute of Astrophysics, FORTH, N. Plastira 100, 70013, Heraklion, Greece\label{forth}}\and 
{Max-Planck-Institut f{\"u}r Radioastronomie, Auf dem H{\"u}gel 69, 53121 Bonn, Germany\label{mpifr}}\and
\ifnum\wm>1{Department of Physics, Indian Institute of Technology Roorkee, Roorkee-247667, India\label{IITR}}\and\fi
\ifnum\wm>1{Department of Electrical Engineering, IIT Hyderabad, Kandi, Telangana 502284, India \label{IITH_El}}\and\fi
\ifnum\wm=5{Cosmology, Universe and Relativity at Louvain (CURL), Institute of Mathematics and Physics, University of Louvain, 2 Chemin du Cyclotron, 1348 Louvain-la-Neuve, Belgium \label{curl}}\and\fi
{Universit{\'e} Paris Cit{\'e}, CNRS, Astroparticule et Cosmologie, 75013 Paris, France\label{apc}}\and
\ifnum\wm>1{The Institute of Mathematical Sciences, C. I. T. Campus, Taramani, Chennai 600113, India \label{IMSc}}\and\fi
\ifnum\wm>1{Homi Bhabha National Institute, Training School Complex, Anushakti Nagar, Mumbai 400094, India \label{HBNI}}\and\fi
{Fakult{\"a}t f{\"u}r Physik, Universit{\"a}t Bielefeld, Postfach 100131, 33501 Bielefeld, Germany\label{unibi}}\and
\ifnum\wm=5{Scuola Internazionale Superiore di Studi Avanzati: Via Bonomea 265, I-34136 Trieste, Italy and INFN Sezione di Trieste \label{sissa}}\and\fi
{ASTRON, Netherlands Institute for Radio Astronomy, Oude Hoogeveensedijk 4, 7991 PD, Dwingeloo, The Netherlands\label{astron}}\and
\ifnum\wm>1{Department of Physical Sciences, Indian Institute of Science Education and Research, Mohali, Punjab 140306, India \label{IISERM}}\and\fi
{Laboratoire de Physique et Chimie de l'Environnement et de l'Espace, Universit\'e d'Orl\'eans / CNRS, 45071 Orl\'eans Cedex 02, France \label{lpc2e}}\and
{Observatoire Radioastronomique de Nan\c{c}ay, Observatoire de Paris, Universit\'e PSL, Université d'Orl\'eans, CNRS, 18330 Nan\c{c}ay, France\label{nancay}}\and
{Dipartimento di Fisica ``G. Occhialini", Universit{\'a} degli Studi di Milano-Bicocca, Piazza della Scienza 3, I-20126 Milano, Italy\label{unimib}}\and
{INFN, Sezione di Milano-Bicocca, Piazza della Scienza 3, I-20126 Milano, Italy\label{infn-unimib}}\and
{INAF - Osservatorio Astronomico di Brera, via Brera 20, I-20121 Milano, Italy\label{inaf-brera}}\and
{Institute for Gravitational Wave Astronomy and School of Physics and Astronomy, University of Birmingham, Edgbaston, Birmingham B15 2TT, UK\label{unibir}}\and
{INAF - Osservatorio Astronomico di Cagliari, via della Scienza 5, 09047 Selargius (CA), Italy\label{inaf-oac}}\and
{Hellenic Open University, School of Science and Technology, 26335 Patras, Greece\label{HOU}}\and
\ifnum\wm=5{Université de Gen\`eve, Département de Physique Théorique and Centre for Astroparticle Physics, 24 quai Ernest-Ansermet, CH-1211 Genéve 4, Switzerland\label{unige}}\and\fi
\ifnum\wm=5{CERN, Theoretical Physics Department, 1 Esplanade des Particules, CH-1211 Gen\'{e}ve 23, Switzerland\label{CERN}}\and\fi
{Kavli Institute for Astronomy and Astrophysics, Peking University, Beijing 100871, P. R. China\label{kiaa}}\and
\ifnum\wm>1{Department of Astronomy and Astrophysics, Tata Institute of Fundamental Research, Homi Bhabha Road, Navy Nagar, Colaba, Mumbai 400005, India \label{TIFR}}\and\fi
\ifnum\wm>1{Department of Physics, IIT Hyderabad, Kandi, Telangana 502284, India \label{IITH_Ph}}\and\fi
\ifnum\wm>1{Department of Physics and Astrophysics, University of Delhi, Delhi 110007, India \label{UoD}}\and\fi
\ifnum\wm>1{Department of Earth and Space Sciences, Indian Institute of Space Science and Technology, Valiamala, Thiruvananthapuram, Kerala 695547,India \label{IIST}}\and\fi
\ifnum\wm=5{School of Mathematics and Physics, Faculty of Engineering and Physical Science, University of Surrey, Guildford GU2 7XH, UK\label{surrey}}\and\fi
{School of Physics, Faculty of Science, University of East Anglia, Norwich NR4 7TJ, UK\label{uea}}\and
\ifnum\wm=5{Sternberg Astronomical Institute, Moscow State University, Universitetsky pr., 13, Moscow 119234, Russia\label{SAI}}\and\fi
{Max Planck Institute for Gravitational Physics (Albert Einstein Institute), Am M{\"u}hlenberg 1, 14476 Potsdam, Germany\label{aei}}\and
{Gran Sasso Science Institute (GSSI), I-67100 L'Aquila, Italy \label{gssi}}\and
{INFN, Laboratori Nazionali del Gran Sasso, I-67100 Assergi, Italy \label{lngs}}\and 
\ifnum\wm>1{National Centre for Radio Astrophysics, Pune University Campus, Pune 411007, India \label{NCRA}}\and\fi
\ifnum\wm>1{Kumamoto University, Graduate School of Science and Technology, Kumamoto, 860-8555, Japan \label{KU_J}}\and\fi
{Universit{\'a} di Cagliari, Dipartimento di Fisica, S.P. Monserrato-Sestu Km 0,700 - 09042 Monserrato (CA), Italy\label{unica}}\and
{Department of Astrophysics/IMAPP, Radboud University Nijmegen, P.O. Box 9010, 6500 GL Nijmegen, The Netherlands\label{imapp}}\and
\ifnum\wm>1{Department of Physical Sciences,Indian Institute of Science Education and Research Kolkata, Mohanpur, 741246, India \label{IISERK}}\and\fi
\ifnum\wm>1{Center of Excellence in Space Sciences India, Indian Institute of Science Education and Research Kolkata, 741246, India \label{CESSI}}\newpage \and \fi
{School of Physics, Trinity College Dublin, College Green, Dublin 2, D02 PN40, Ireland\label{tcd}}\and
{Jodrell Bank Centre for Astrophysics, Department of Physics and Astronomy, University of Manchester, Manchester M13 9PL, UK\label{jbca}}\and
\ifnum\wm>1{Department of Physics, St. Xavier’s College (Autonomous), Mumbai 400001, India \label{XCM}}\and\fi
\ifnum\wm>1{Department of Astronomy,School of Physics, Peking University, Beijing 100871, P. R. China\label{pku}}\and\fi
\ifnum\wm>1{National Astronomical Observatories, Chinese Academy of Sciences, Beijing 100101, P. R. China\label{naoc}}\and\fi
{E.A. Milne Centre for Astrophysics, University of Hull, Cottingham Road, Kingston-upon-Hull, HU6 7RX, UK\label{milne}}\and
{Centre of Excellence for Data Science, Artificial Intelligence and Modelling (DAIM), University of Hull, Cottingham Road, Kingston-upon-Hull, HU6 7RX, UK\label{daim}}\and
\ifnum\wm=5{Laboratory of Astrophysics, \'Ecole Polytechnique F\'ed\'erale de Lausanne, CH-1015 Lausanne, Switzerland\label{EPFL}}\and\fi
\ifnum\wm>1{Department of Physics, BITS Pilani Hyderabad Campus, Hyderabad 500078, Telangana, India \label{BITS}}\and\fi
\ifnum\wm>1{Joint Astronomy Programme, Indian Institute of Science, Bengaluru, Karnataka, 560012, India \label{IISc}}\and\fi
{Arecibo Observatory, HC3 Box 53995, Arecibo, PR 00612, USA\label{arecibo}}\and
{IRFU, CEA, Université Paris-Saclay, F-91191 Gif-sur-Yvette, France \label{irfu}}\and
\ifnum\wm>1{Raman Research Institute India, Bengaluru, Karnataka, 560080, India \label{RRI}}\and\fi
{Institut f\"{u}r Physik und Astronomie, Universit\"{a}t Potsdam, Haus 28, Karl-Liebknecht-Str. 24/25, 14476, Potsdam, Germany\label{uni-potsdam}}\and
\ifnum\wm=5{Kazan Federal University, 18 Kremlyovskaya, 420008 Kazan, Russia\label{KFU}}\and\fi
\ifnum\wm>1{Department of Physics, IISER Bhopal, Bhopal Bypass Road, Bhauri, Bhopal 462066, Madhya Pradesh, India \label{IISERB}}\and\fi
{Ollscoil na Gaillimhe --- University of Galway, University Road, Galway, H91 TK33, Ireland\label{uog}}\and
\ifnum\wm>1{Center for Gravitation, Cosmology, and Astrophysics, University of Wisconsin-Milwaukee, Milwaukee, WI 53211, USA \label{CGCA}}\and\fi
\ifnum\wm>1{Division of Natural Science, Faculty of Advanced Science and Technology, Kumamoto University, 2-39-1 Kurokami, Kumamoto 860-8555, Japan \label{KU_J1}}\and\fi
\ifnum\wm>1{International Research Organization for Advanced Science and Technology, Kumamoto University, 2-39-1 Kurokami, Kumamoto 860-8555, Japan \label{KU_J2}}\and\fi
{Laboratoire Univers et Th{\'e}ories LUTh, Observatoire de Paris, Universit{\'e} PSL, CNRS, Universit{\'e} de Paris, 92190 Meudon, France\label{luth}}\and
\ifnum\wm=5{Max Planck Institute for Gravitational Physics (Albert Einstein Institute), Leibniz Universit\"at Hannover, Callinstrasse 38, D-30167, Hannover, Germany\label{aei2}}\and\fi
{Florida Space Institute, University of Central Florida, 12354 Research Parkway, Partnership 1 Building, Suite 214, Orlando, 32826-0650, FL, USA\label{FSI}}\and
{Ruhr University Bochum, Faculty of Physics and Astronomy, Astronomical Institute (AIRUB), 44780 Bochum, Germany \label{airub}}\and
{Advanced Institute of Natural Sciences, Beijing Normal University, Zhuhai 519087, China \label{bnuz}}
}

   \date{Received May 8, 2023; accepted }
\titlerunning{GWB Search}
\authorrunning{EPTA: CGW search}

 
  \abstract
    {
        We present the results of a search for continuous gravitational wave signals (CGWs) in the second data release (DR2) of the European Pulsar Timing Array (EPTA) collaboration. The most significant candidate event from this search has a gravitational wave frequency of 4-5 nHz. Such a signal could be generated by a supermassive black hole binary (SMBHB) in the local Universe. We present the results of a follow-up analysis of this candidate using both Bayesian and frequentist methods. The Bayesian analysis gives a Bayes factor of 4 in favor of the presence of the CGW over a common uncorrelated noise process. In contrast, the frequentist analysis estimates the p-value of the candidate to be $< 1\%$, also assuming the presence of common uncorrelated red noise. However, comparing a model that includes both a CGW and a gravitational wave background (GWB) to a GWB only, the Bayes factor in favor of the CGW model is only $0.7$. Therefore, we cannot conclusively determine the origin of the observed feature, but we cannot rule it out as a CGW source. We present results of simulations that demonstrate that data containing a weak gravitational wave background can be misinterpreted as data including a CGW and vice versa, providing two plausible explanations of the EPTA DR2 data. Further investigations combining data from all PTA collaborations will be needed to reveal the true origin of this feature.    
    }

   \maketitle
%

\section{Introduction}
The population of SMBHBs in the relatively local Universe is the most promising astrophysical source of gravitational waves (GWs) at nanohertz frequencies, probed by pulsar timing array (PTA) observations. The signal is generated by binaries in wide orbits with periods of months to years. Each binary is far from merger and evolving slowly, so the emitted GWs are almost monochromatic. However, the incoherent superposition of GWs from many binaries creates a stochastic GW background (SGWB) signal with a characteristic broad red-noise type spectrum. A search of the second data release (DR2) of the European Pulsar Timing Array (EPTA) for an SGWB was reported in~\citep{wm3}. This analysis reported increasing evidence for an SGWB, based on seeing a red noise process with a common spectral shape in all pulsars and seeing evidence that the correlation of the signal between pairs of pulsars was consistent with the forecasted Hellings-Downs (HD) correlation curve that is expected from an SGWB. The statistical significance reported in~\citep{wm3} is not yet high enough to claim a detection, but the data is starting to show some evidence for an SGWB. 

The EPTA DR2 includes 25 pulsars selected to optimize for detection of the HD correlations, based on the methods described in \cite{2023MNRAS.518.1802S}.  The analyzed data were collected with six EPTA telescopes: the Effelsberg Radio Telescope in Germany, the Lovell Telescope in the UK, the Nan\c{c}ay Radio Telescope in France, the Westerbork Synthesis Radio Telescope in the Netherlands, the Sardinia Telescope in Italy and the Large European Array for Pulsars. In this paper, we have used the \texttt{DR2new} subset of the full data~\cite {wm3}. It includes only the last 10.3 years of data, which was collected with new-generation wide-band backends.

The analysis of the EPTA data DR2new has reported evidence of an HD correlated background as opposed to a common red noise process ~\cite{wm3}. However, this evidence may have resulted 
from a stochastic GW signal or from one or a few bright continuous gravitational wave (CGW) sources in the data (e.g. see \citep{Allen_2023} for details). Therefore, it is important to determine whether such CGW sources are present. This paper aims to search for individual CGW sources in the EPTA data DR2new.

This paper uses frequentist and Bayesian approaches to search for a CGW source. We adopt the model of a single binary system in a circular orbit. We analyze the data using both Earth-term only and a full signal (Earth + pulsar terms) model. For each pulsar, we assume the custom-made noise model reported in \citep{wm2}. We also allow for the presence of a common red noise (CRN) component in the data. Evidence for a CRN was reported in the analysis of the reduced EPTA DR2 dataset comprising the six pulsars with the best timing accuracy \cite{ccg+21}. The 6-pulsar dataset was not informative on the nature of this CRN signal, but the more recent 25-pulsar analysis reported in~\citep{wm3} favors an SGWB origin for this noise. In this analysis, we consider models that include a deterministic CGW signal and one of three different noise models: individual pulsar noises only (PSRN), PSRN plus a common \textit{uncorrelated} red noise (CURN) process, or PSRN plus an SGWB with Hellings-Downs (GWB) correlations between the pulsars. Simple power-law power spectral densities will represent all common noises.  

We have conducted a Bayesian search for CGWs across a wide frequency band by splitting the dataset into sub-bands of width $\Delta \log_{10}f_{gw} = 0.05$. We follow up on the most significant candidate from this search with a detailed analysis. In addition, we have performed an analysis on simulated datasets generated with noise properties consistent with the posterior distribution inferred from the actual data.vThe evaluated Bayes factors for an SGWB model versus SGWB with the addition of a CGW are close to unity. Therefore, the evidence provided by the data is equally consistent with both the inclusion or the absence of a CGW source. In other words, there is no preference for one hypothesis over the other based on the data, even though the CGW model has 58 additional parameters. This indicates that more data is needed to make a conclusive decision.

The paper is organized as follows. In Section~\ref{sec:method}, we describe the model used to describe the data and the frequentist and Bayesian methods we employ in our analysis. In Section~\ref{sec:results}, we present the results of the analysis of the EPTA DR2 dataset. In Section~\ref{sec:simulations}, we discuss results from a simulation study that we undertook to understand the results of the EPTA DR2 analysis. Finally, in Section~\ref{sec:conclusions}, we summarize our results and current findings.

\section{Methods}
\label{sec:method}

\subsection{Noise model}

We adopt the model for the noise in a single pulsar described in \citep{wm2}, in which timing residuals are written as
\begin{equation}
	\delta t = \underbrace{
\textrm{M} \epsilon + n_{\rm{WN}} + n_{\rm{RN}} + n_{\rm{DM}} + n_{\rm{Sv}}}_{\rm PSRN} + \,\underbrace{n_{\rm{CRN}}}_{\rm Common \, Red \, Noise}  + \underbrace{s}_{\rm CGW}.
\end{equation}

The timing model error, $\textrm{M} \epsilon$, is represented by a linear model based on the design matrix $\textrm{M}$ and an offset from the nominal timing model parameters, $\epsilon$. The white noise component $n_{\rm{WN}}$ is described by two parameters for each backend, which apply multiplicative (EFAC) and additive (EQUAD) corrections to the estimated timing uncertainty. The pulsar red noise, $n_{\rm{RN}}$, dispersion measure variations,  $n_{\rm{DM}}$, and scattering variations, $n_{\rm{Sv}}$, are each represented by an incomplete Fourier basis defined at $i/T_{\rm{obs}}$ frequency bins ($i$ is integer). The amplitudes are assumed to be generated by a stationary Gaussian process~\citep{vv2014}, with PSD described by a power-law, characterized by spectral index, $\gamma$, and the amplitude, $A$, at reference frequency $f_{\rm{ref}}$=1/year. The noise models, including the number of frequencies in the Fourier basis, are customized for each pulsar, as described in~\cite{wm2}.  We call the model that includes all of the aforementioned noise components the PulSaR Noise (PSRN) model. 

We also allow for the presence of a common red noise (CRN), $n_{\rm CRN}$, affecting all the pulsars, that can take the form of an uncorrelated noise among pulsars (CURN) or a gravitational wave background (GWB) with a correlation described by the HD curve. We model the properties of the CRN similarly to the individual pulsar red noises, using an incomplete Fourier basis, with amplitudes described by a Gaussian process with a power-law PSD. In the Bayesian analysis below, we have used either three or nine Fourier bins for describing the CRN, and we have adopted the same priors for the pulsar noise components as presented in \cite{wm3}. We refer the reader to \cite{wm2, 2022MNRAS.509.5538C} for a more complete description of the noise models.

The final component of the model for the timing residuals is a continuous gravitational wave (CGW) signal, $s$. This will be described in the next section.

\subsection{Continuous gravitational wave model}
\label{sec:CGWmodel}

A supermassive black hole binary (SMBHB) system in a circular orbit produces monochromatic and quasi-non-evolving GWs \citep{arzoumanian2023nanograv, Falxa_2023}. Such signals induce pulsar timing residuals $s_a(t, \hat{\Omega})$ of the form :
\begin{equation}
	\label{eqn:resid}
	s_a(t, \hat{\Omega}) = \sum _{A=+,\times} F^A(\hat{\Omega})[s_A (t) - s_A (t - \tau_a)],
\end{equation}
where $s_A(t)$ and $s_A(t-\tau_a)$ are referred to as the \textit{Earth term} (ET) and the \textit{Pulsar term} (PT), $F^A(\hat{\Omega})$ are the antenna pattern functions that characterize how each GW polarisation, $+,\times$, affects the residuals as a function of the sky location of the source, $\hat{\Omega}$ is the direction of propagation of the GW and $\tau_a$ is a delay time between the source and pulsar $a$. The full expressions for $s_A(t)$ are:
\begin{eqnarray}
		s_+(t) = \frac{\mathcal{M}^{5/3}}{d_L \omega(t)^{1/3}} &\left\{ -\sin\left[2\Phi (t)\right](1+\cos^2 \iota) \cos 2\psi \right. \nonumber \\
		&\left.- 2\cos\left[2\Phi(t)\right]\cos \iota \sin 2\psi  \vphantom{(\cos^2 \iota)} \right\}  ,\\ 
		s_\times(t) = \frac{\mathcal{M}^{5/3}}{d_L \omega(t)^{1/3}} &\left\{ -\sin\left[2\Phi (t)\right](1+\cos^2 \iota) \cos 2\psi \right. \nonumber \\
		&\left. + 2\cos\left[2\Phi(t)\right]\cos \iota \sin 2\psi  \vphantom{(\cos^2 \iota)} \right\} ,
\end{eqnarray}
with $\mathcal{M}$ the chirp mass, $d_L$ the luminosity distance to the source, $\omega(t) = \pi f_{gw}(t)$ the time dependent frequency of the GW, $\iota$ the inclination, $\psi$ the polarisation angle and $\Phi(t)$ the time dependent phase of the GW. The amplitude of the GW is given by:
\begin{equation}\label{eq:h_mc_f}
    h = 2\frac{\mathcal{M}^{5/3}}{d_L} (\pi f_{gw})^{2/3} \, .
\end{equation}

For a slowly evolving binary, $\omega(t)$ is considered constant ($\omega(t)=\omega_0$) over the duration of PTA observations of $\sim$10 years, giving for the Earth and Pulsar phases:
\begin{eqnarray}
    \Phi(t) & = & \Phi_0 + \omega_0 t, \\
    \Phi(t-\tau_a) & = & \Phi_0 + \Phi_a + \omega(t-\tau_a) t.
\end{eqnarray}

Nonetheless, the difference in frequency between Earth and Pulsar terms can be significant. The frequency of the pulsar term can be computed using the leading order radiation reaction evolution:
\begin{equation}
    \omega(t) = \omega _0 \bigg [ 1 - \frac{256}{5} \mathcal{M}^{5/3} \omega_0 ^{8/3} (t - t_0) \bigg]^{-3/8}.
\end{equation}

This difference in $\omega(t)$ is determined by the time delay $\tau_a$ given by:
\begin{equation}
	\tau _a = L_a (1 + \hat{\Omega} \cdot \hat{p}_a),
\label{eq:psr_time_delay}
\end{equation}
where $L_a$ the distance between the  Earth and pulsar $a$ and $\hat{p}_a$ is a unit vector pointing to pulsar $a$. If the SMBHB has significantly evolved during the time $\tau_a$, the Earth term will have a higher frequency than the Pulsar term. This will usually be the case for frequencies above $\sim$10 nHz. For binaries at lower frequencies, binary evolution is typically negligible, and both terms will have the same frequency (within the resolution of the PTA) but different phases. The characterisation of the pulsar term can be difficult because the distance $L_a$ is known with poor accuracy. Consequently, the pulsar distance $L_a$ and phase $\Phi_a$ must be treated as free parameters that are fitted while searching for the signal \citep{corbin_cornish}. In our analysis, we use a Gaussian prior on the distances $L_a$ with the measured mean, $\mu_a$, and uncertainty, $\sigma_a$, from \cite{Verbiest_2012}\footnote{\url{https://www.atnf.csiro.au/research/pulsar/psrcat/}}. For the pulsars not included in that paper, we use a mean of 1 kpc and an error of 20\%.

\begin{table}
\centering
\begin{tabular}{c|c|c}
CGW parameter & Prior & Range \\
\hline
$\log_{10} h$ & Uniform & [-18, -11] \\
$\log_{10} f_{gw}$ & Uniform & [-9, -7.85] \\
$\log_{10} \mathcal{M}$ & Uniform & [7, 11] \\
$\Phi_0$ & Uniform & [0, $\pi$] \\
$\cos \iota$ & Uniform & [-1, 1] \\
$\psi$ & Uniform & [0, $\pi$] \\
$\cos \theta$ & Uniform & [-1, 1] \\
$\phi$ & Uniform & [0, $2\pi$] \\
$\Phi_a$ & Uniform & [0, $\pi$] \\
$L_a$ & Normal, $\mathcal{N}(\mu_a,\sigma_a)$ & [-$\infty$, $\infty$] \\
\end{tabular}
\caption{List of parameters of the continuous gravitational wave model with their respective priors and ranges.}
\label{tab:cw_params}
\end{table}

\subsection{Frequentist analysis}
\label{sec:FreqAn}
We analyse the data using a frequentist approach based on the Earth term-only  $\mathcal{F}_e$-statistic (\cite{Babak:2011mr, Ellis:2012zv}). The $\mathcal{F}_e$ detection statistic is the log-likelihood maximized over the ``extrinsic'' CGW parameters ($h$, $\iota$, $\psi$, $\phi_0$) for a fixed set of intrinsic parameters ($\theta$, $\phi$, $f_{gw}$).  
If the residuals are Gaussian, the null distribution is expected to be a $\chi^2$ distribution with 4 degrees of freedom. In the presence of a signal, $\mathcal{F}_e$ is distributed as a non-central $\chi^2$-distribution with non-centrality parameter related to the square of the signal-to-noise-ratio $(s(t)|s(t))$\footnote{$(s|s)$ denotes the noise weighted inner product, $(s|s) = s C^{-1} s^\intercal$, with $C$ the covariance matrix of the noise model.} (see~\cite{Ellis:2012zv} for further details). However, to calculate $\mathcal{F}_e$, we need to make assumptions about the noise properties. We take two different approaches: $(i)$ we use the posterior distributions obtained from fitting the noise parameters to obtain a distribution of $\mathcal{F}_e$ for each set of intrinsic parameters; $(ii)$ we fix the noise parameters to their 
maximum likelihood estimates, as is often done for the EFAC and EQUAD parameters.
The second approach is standard within frequentist analysis, but we also use (i) for the red noise components to emphasize that the inferred parameters have rather significant uncertainties. Varying the noise parameters generates a  distribution of the optimal statistic for each choice of intrinsic parameters and thus brings an element of Bayesian approaches into this frequentist analysis.

We want to evaluate the significance of the highest $\mathcal{F}_e$ measured on the observed data by computing the p-value, which is a statement about how improbable it would be to draw the observed data if no signal was present. Calculating the true p-value requires the true distribution of $\mathcal{F}_e$ in the absence of signal (the null distribution), which we do not have. Building a proper null distribution for the $\mathcal{F}_e$ statistics preserving the properties of the noise (of observed data) requires an extensive research program and is outside the scope of this paper. Instead, we propose two simplified ways of estimating the significance:
(i) using the theoretical null distribution of 2$\mathcal{F}_e$, 
assuming that our noise model is correct and complete. The theoretical distribution behaves as a $\chi^2$ with 4 degrees of freedom, (ii) shuffling the pulsars on the sky but keeping the "source" position fixed to its maximum likelihood value. 
The second approach requires further explanations. The signal-to-noise ratio for the candidate event is dominated by several pulsars, which implies that a simple shuffle of the pulsars will not destroy a monochromatic GW but will change its sky position and extrinsic parameters (initial phase, inclination, etc). By shuffling the pulsars on the sky, we ask, ``how likely to obtain the observed value of $\mathcal{F}_e$ by random position of pulsars''. This approach has the advantage of making no distributional assumptions about the pulsars' noise properties. Further, it requires no minimal match or any orthogonality condition between shuffled pulsar positions (see discussion on this topic in \cite{tlb+2017, dimarco2023robust})
The sky shuffled distribution was obtained as follows:

\begin{itemize}
    \item i) We produce 3000 shuffles by drawing random pulsar positions uniformly in the sky.  
    \item ii) For each of the 3000 draws, we evaluate $\mathcal{F}_e$ for 1000 noise parameters drawn from the posterior distributions obtained in \citep{wm2} and take the mean value.
    \item iii) We produce a histogram of the 3000 $\mathcal{F}_e$ statistic values, representing the background distribution.
    \item iv) We repeat (i-iii) 20 times, obtaining a slightly different histogram each time due to differences in the 3000 shuffles and the noise realizations. This allows us to estimate the uncertainty in the background distribution and, therefore, in the computed p-value.
\end{itemize}

We will apply this procedure to the CGW candidate in  Section~\ref{sec:results}. 

\subsection{Bayesian analysis}
We also conducted a Bayesian analysis to obtain posterior probability distributions for the noise and signal parameters in the model described in Section~\ref{sec:method}. We make use of MCMC samplers \texttt{PTMCMC} \citep{ptmcmc}), \texttt{QuickCW} \citep{Becsy:2022zbu} and \texttt{Eryn} \citep{Karnesis:2023ras} to explore the parameter space. The Bayesian analysis allows us to perform parameter inference and model selection. The latter is quantified by evaluating the Bayes factor: the ratio of the marginal posterior distributions (or evidence) for two different models. The marginal posterior is a quantity challenging to compute and can be estimated numerically using parallel tempering or nested sampling (\cite{ski2004}).

In this paper, we use \texttt{ENTERPRISE} \citep{enterprise, enterprise_extensions} to evaluate the posterior probability for a given model. We compute the Bayes factors using the product-space method (\cite{hhh+2016}) implemented in \texttt{ENTERPRISE} and through Reversible Jump Markov Chain Monte Carlo (RJMCMC), as implemented in \texttt{Eryn}. In both approaches, at each step of the Markov Chain, either the parameters within the current model can be updated, or a switch to a different model can be proposed. The acceptance rule for the model switch is defined to ensure that detailed balance is maintained, thus ensuring that the stationary distribution of the Markov Chain is the desired posterior distribution over models. The sampler will spend more time exploring the model with the highest marginal posterior probability. The Bayes factor $\mathcal{B}^{A/B}$ between models $A$ and $B$ can then be calculated as the ratio between the final number of chain samples corresponding to each model.

In the product-space method, the chain samples are in a hypermodel space, which is a union of all the parameters of all the models being considered. An additional parameter determines which model is active within each sample, while inactive parameters undergo a random walk during the within-model steps. The effect is that the product space method retains some memory of where it had been exploring the other models, which can increase the probability that a proposed switch back to the other model is accepted. In RJMCMC, the chain typically only samples in the parameters of the currently active model and does not retain any memory. This can lead to lower model-switch acceptance rates but guarantees a more complete exploration of the parameter spaces of the different models. Note that RJMCMC methods applied to PTA data analysis have been previously explored in \cite{Becsy:2019dim}.

\section{Results of data analysis}
\label{sec:results}
\subsection{Frequentist analysis}
\label{subsec:fstat_results}

Within the frequentist approach, we want to maximize the detection statistic ($\mathcal{F}_e$ in our case) over all intrinsic parameters of the model. We perform the search using the noise models described in Section~\ref{sec:method} for 100 logarithmically spaced frequencies from 1 to 100 nHz dividing the sky into 3072 different pixels using \texttt{healpix} (\cite{Zonca2019, 2005ApJ...622..759G}. 

To account for the fact that the noise model has broad posteriors, we use the posterior samples of the noise parameters obtained in \citep{wm2} to calculate $\mathcal{F}_e$ for fixed CGW parameters and average $\mathcal{F}_e$ over 1000 randomly drawn samples of the PSRN model. The maximum of $\mathcal{F}_e$ is found at 4.64 nHz, consistent with the results of the Bayesian analyses described in \autoref{sec:bayesian}.

\begin{figure}
\centering
\includegraphics[width=0.98\linewidth]{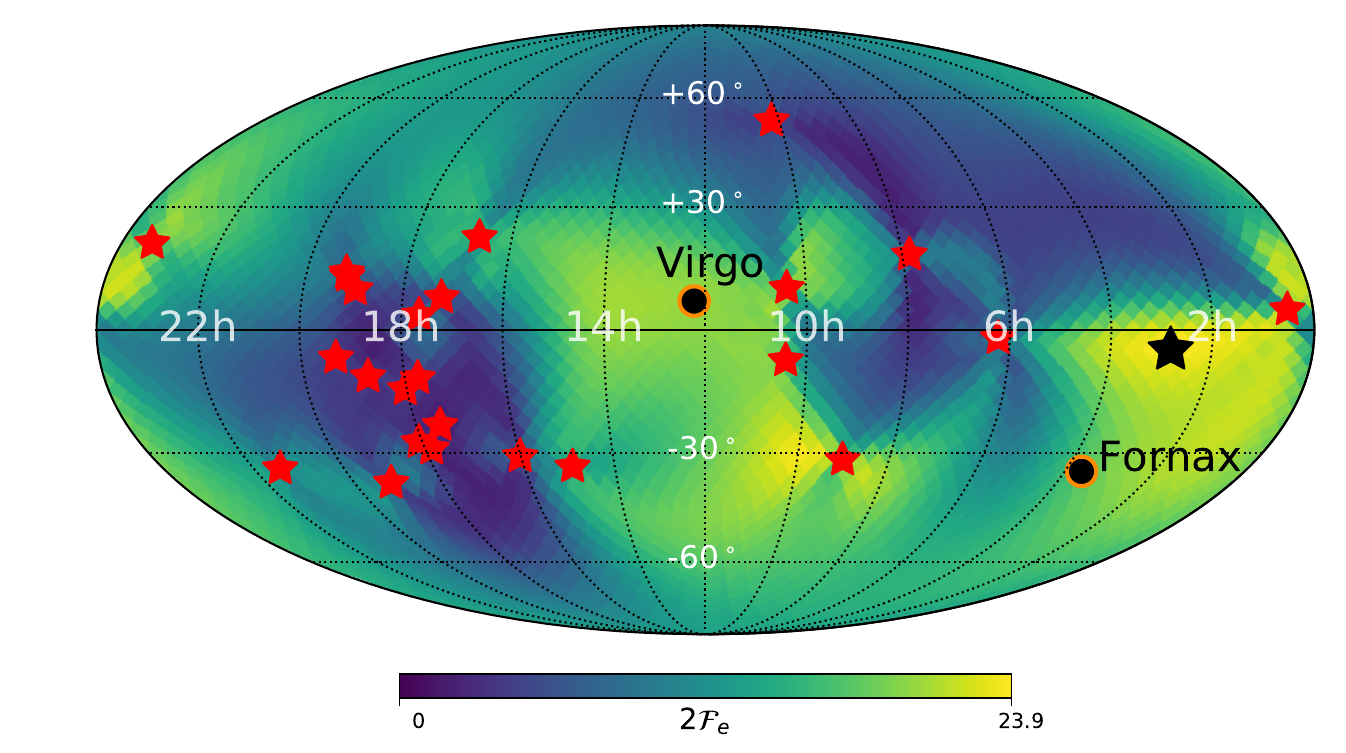}
\caption{$\mathcal{F}_e$-statistic of the candidate source at $f_{gw}=4.64$~nHz averaged over the noise uncertainties for the custom PSRN model. The black star shows the position of highest $\mathcal{F}_e$, whereas the red stars show the positions of the pulsars. The Fornax and Virgo clusters are shown as black dots.}
\label{fig:Festat_map}
\end{figure}

The sky distribution of $\mathcal{F}_e$ at this frequency is given in \autoref{fig:Festat_map}. The region of high statistic value (bright yellow) is relatively sparse and inconclusive with regard to the localisation of the CGW candidate.  The maximum $\mathcal{F}_e$ is depicted by a black star and corresponds to a region of the sky where we lack pulsars and, hence where the array is expected to be less sensitive.

\begin{figure}
\centering
\includegraphics[width=0.98\linewidth]{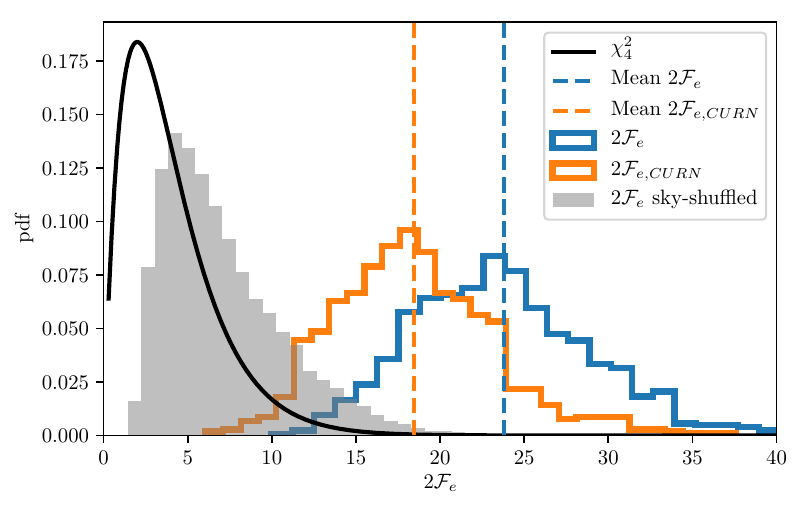}
\caption{Distribution of $\mathcal{F}_e$-statistic over the noise uncertainties without CURN (blue) and with CURN (orange) at 4.64 nHz. The null distributions of the $\mathcal{F}_e$ are obtained from the analysis of the EPTA \texttt{DR2new} data with shuffled sky positions (grey shaded region) and from the theoretical formula of a $\chi^2_4$-distribution (black solid line).}
\label{fig:Festat_CURN_no_CURN}
\end{figure}

The analysis was repeated by including the CURN component in the noise model, and results are presented in \autoref{fig:Festat_CURN_no_CURN}. We show two distributions of $\mathcal{F}_e$. These are both evaluated at the optimal sky position and at the GW frequency 4.64 nHz and are obtained by varying the noise parameters (random draw) with (orange histogram) and without (blue histogram) the CURN component. Inclusion of the CURN slightly reduces the significance of the CGW candidate.

To evaluate the p-value, we compute the distribution of sky-shuffled pulsars according to the steps outlined in \autoref{sec:FreqAn} at the 
CGW candidate parameter values (maximising the noise averaged $\mathcal{F}_e$). These results are indicated by the grey shaded distribution in \autoref{fig:Festat_CURN_no_CURN}. The theoretical $\chi^2$ null distribution is shown as a black curve. The sky-shuffled distribution of $2\mathcal{F}_e$ is close to the theoretical $\chi^2_4$ distribution but not identical. This could be due to (i) non-gaussian noise present in the array; (ii) we are asking different questions in two approaches.

\begin{table}
\caption{
Statistical significance of the candidate source at 4.64 nHz. The p-values for the $\chi^2_4$ are obtained using the maximum likelihood noise parameters and the pulsar sky-shuffled p-value from the mean of the $2\mathcal{F}_e$ distribution. We show the p-values for the custom pulsar noise PSRN, $p(\mathcal{F}_e)$, or also including a common uncorrelated red noise CURN+PSRN, $p(\mathcal{F}_{e,\rm{CURN}})$.
}
\def\arraystretch{1.5} 
\centering
\begin{tabular}{c||c|c} 
 \hline
    & $p(\mathcal{F}_e)$ & 
    $p(\mathcal{F}_{e,\rm{CURN}})$ \\
 \hline \hline
$\chi ^2 _4$ &	$9 \times 10^{-5}$ & $1 \times 10^{-3}$ \\
 Sky scrambles & $(2 \pm 1) \times 10^{-4}$ & $(8 \pm 0.4) \times 10^{-3}$ \\
 \hline
\end{tabular}
\label{tab:fstat_sigmas}
\end{table}

We compute p-values using the obtained distributions and the measured mean values of the orange and blue $\mathcal{F}_e$ distributions. The results are summarized in \autoref{tab:fstat_sigmas}, the top row is for the theoretical distribution and the second row is for the sky-shuffled distribution (with uncertainty). The obtained p-value for $\mathcal{F}_e$ corresponds to about $3.5\sigma$ while $\mathcal{F}_{e,CURN}$ corresponds to about $2.5\sigma$.

In order to apply the full detection pipeline, we would need to set the detection threshold based on the chosen false alarm probability. This, in turn, would require estimating the number of independent cells on the sky-frequency parameter space where the $\mathcal{F}_e$ was evaluated \citep{bps+2016, Ellis:2012zv}. Here, we decided to use the $\mathcal{F}_e$ only to identify potential candidates and produce sky maps of significance without a precise statement of detection within the frequentist approach.

\paragraph{\textbf{Estimating the p-value using $\mathcal{F}_e$ :}}

\begin{figure}
	\centering
	\includegraphics[width=0.98\linewidth]{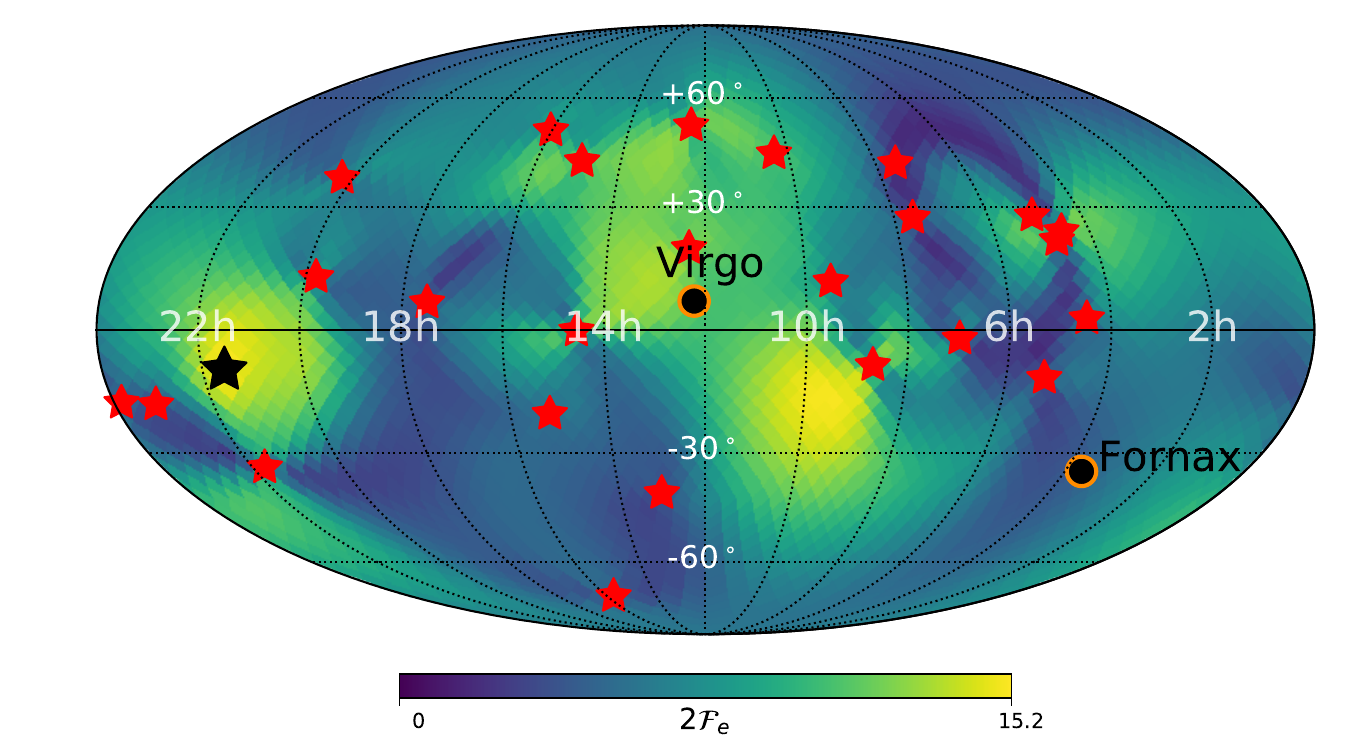}
	\caption{Pulsar-shuffled version of the $\mathcal{F}_e$-statistic of the candidate source at $f_{gw}=4.64$~nHz averaged over the noise uncertainties for the custom PSRN model. The black star shows the position of highest $\mathcal{F}_e$, whereas the red stars show the positions of the pulsars. We see that the highest $\mathcal{F}_e$ pixel changed position but is still significant. The Fornax and Virgo clusters are shown as black dots.}
	\label{fig:Festat_map_scramb}
\end{figure}

In the procedure that we describe, we select the pixel with highest $\mathcal{F}_e$, which is the same as maximising over sky location. Hence, the correct way to produce the background distribution would be to also select the highest $\mathcal{F}_e$ pixel for each scramble. But when doing so, we always find a high $\mathcal{F}_e$ value. This is because the SNR of the observed signal is dominated by a few pulsars ($\sim$3 pulsars). One needs at least 3 pulsars for computing $\mathcal{F}_e$, based on a simple counting of the measurements and unknowns. Any additional pulsar in the array will improve SNR and the source's sky localization (ruling out some parts of the sky). For this reason, shuffling the pulsar position around does not help; we always have a solution based on the 3 best pulsars, and the overall value will depend on the relative contribution of other pulsars in the array. This is not the case when all pulsars are equally good timers as demonstrated in \cite{Babak:2011mr}. In \autoref{fig:Festat_map_scramb} we demonstrated, that sky scramble of the pulsar position still delivers a significant maximum $\mathcal{F}_e$. Note that it does not make $\mathcal{F}_e$ useless, it still indicates a presence of a signal, it rather indicates that the sky scramble is not an appropriate mechanism to obtain the ``null'' dataset.

The reported p-value answers the question: \textrm{ how likely to get an observed $\mathcal{F}_e$ at given point in the parameter space assuming the Gaussian noise and a random sky position (shuffling) of pulsars in the array?} That said, the produced distribution is indeed not the real null distribution. In the text, we used the word \textit{sky-shuffled} instead of \textit{scramble} to avoid confusion with previous works. The p-values appearing in \autoref{tab:fstat_sigmas} report the highest possible significance that is achievable using the real positions of pulsars. What we are essentially doing is searching for mis-modelled CGW-like features, but this procedure does not settle whether this feature is only due to noise.

\subsection{Bayesian analysis}
\label{sec:bayesian}

\begin{figure}
    \centering
    \includegraphics[width=\linewidth]{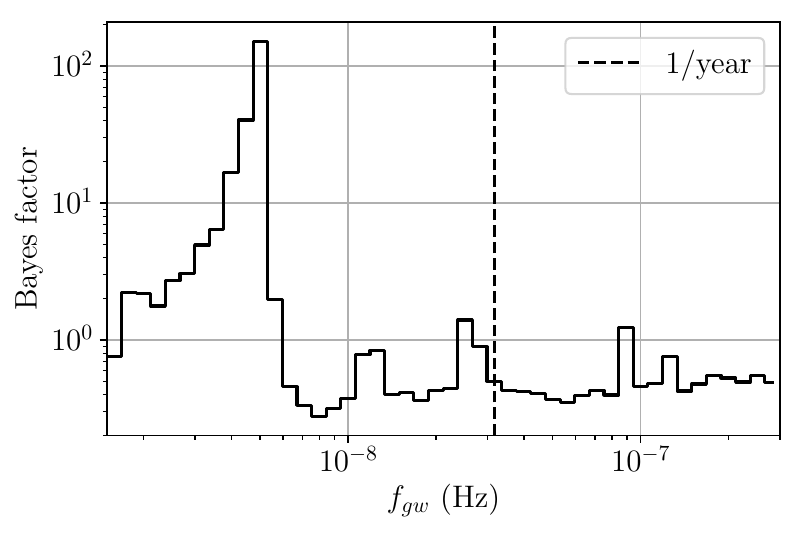}
    \caption{Bayes factor for the model comparison PSRN+CURN+CGW (Earth term) vs PSRN+CURN for 50 logarithmically spaced frequency sub-bands in the region $f_{gw} \in[1.5, 320]$ nHz.}
    \label{fig:BF_search}
\end{figure}

\begin{figure*}
\centering
\includegraphics[width=\linewidth]{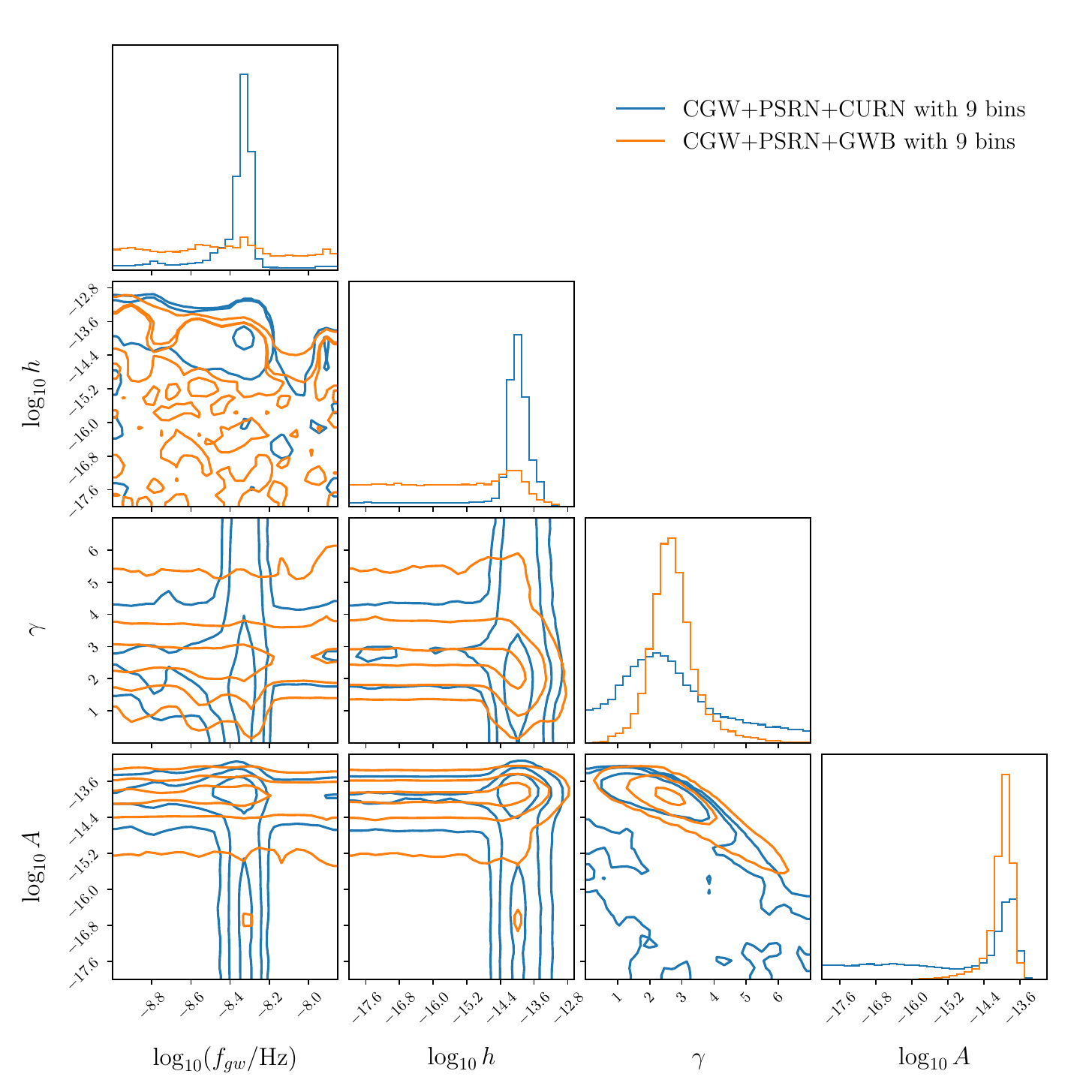}
\caption{Posterior distributions of the CGW search in the second data release of the EPTA \texttt{DR2new}. The posteriors are obtained using a CGW model with Earth and pulsar term, the custom PSRN model and a CRN with either CURN or HD (GWB) correlations represented by 9 frequency bins. We show the posterior distribution for the gravitational wave frequency and amplitude $f_{gw}$, $h$ of the CGW, and the common noise spectral index and amplitude $\gamma$ and $A$. The contours indicate the 1,2,3-$\sigma$ Gaussian contours. }
\label{fig:hd_curn_posterior}
\end{figure*}

We perform a Bayesian CGW search by splitting the frequency range $10^{-9}-10^{-6.5}$ Hz into 50 logarithmically spaced subsegments and assuming Earth term only. We have computed the Bayes factor in each sub-band using \texttt{PTMCMC} and the product-space method. The results are presented in \autoref{fig:BF_search} for the noise model described in Section~\ref{sec:method}. We find a Bayes factor above 100 around 4-5 nHz, and perform a model comparison analysis in a restricted frequency range $f_{gw}\in [10^{-9}, 10^{-7.85}]$ with the priors defined in \autoref{tab:cw_params}. We use Bayes factors as the decision-maker between models. We use \texttt{Eryn} \citep{Karnesis:2023ras} as our fiducial sampler and we crosscheck the results using \texttt{PTMCMC} sampler \citep{ptmcmc}) and \texttt{QuickCW} \citep{Becsy:2022zbu}. The computation of the Bayes factors is performed using RJMCMC and confirmed with the product-space method.
Our findings are summarized in \autoref{tab:main_bayes}, and in the following, we comment on them.
For all models with CGW described below and quoted in the table, we have assumed a circular binary described in ~\autoref{sec:CGWmodel} with Earth and pulsar term, using the priors given in \autoref{tab:cw_params} unless otherwise stated. The validity of these assumptions is discussed at the end of the section.

\begin{table*}
\caption{
Bayes factors obtained for different model comparisons indicate how much the data favor the inclusion of a continuous gravitational wave (CGW) with Earth and pulsar term. The first comparison considers adding a continuous gravitational wave to the pulsar custom noise (PSRN). The second and third model comparisons include an additional common uncorrelated red noise CURN modeled with a power-law with 3 or 9 frequency bins. The fourth and fifth model comparisons include a gravitational wave background GWB modeled with a power-law with 3 or 9 frequency bins.
}
\def\arraystretch{1.5}
\centering
\begin{tabular}{c|c} 
 \hline
   {Model comparison} & 
    {Bayes factor} \\
 \hline
 CGW+PSRN vs PSRN $\qquad \, \, \,$    &   4000		\\
 CGW+PSRN+CURN vs PSRN+CURN, 3 bins     &	12  \\
 CGW+PSRN+CURN vs PSRN+CURN, 9 bins     &	4  \\
 CGW+PSRN+GWB vs PSRN+GWB, 3 bins   &	1	 \\
 CGW+PSRN+GWB vs PSRN+GWB, 9 bins   &	0.7 \\
 \hline
\end{tabular}
\label{tab:main_bayes}
\end{table*}

\begin{figure*}
\centering
\includegraphics[width=0.9\linewidth]{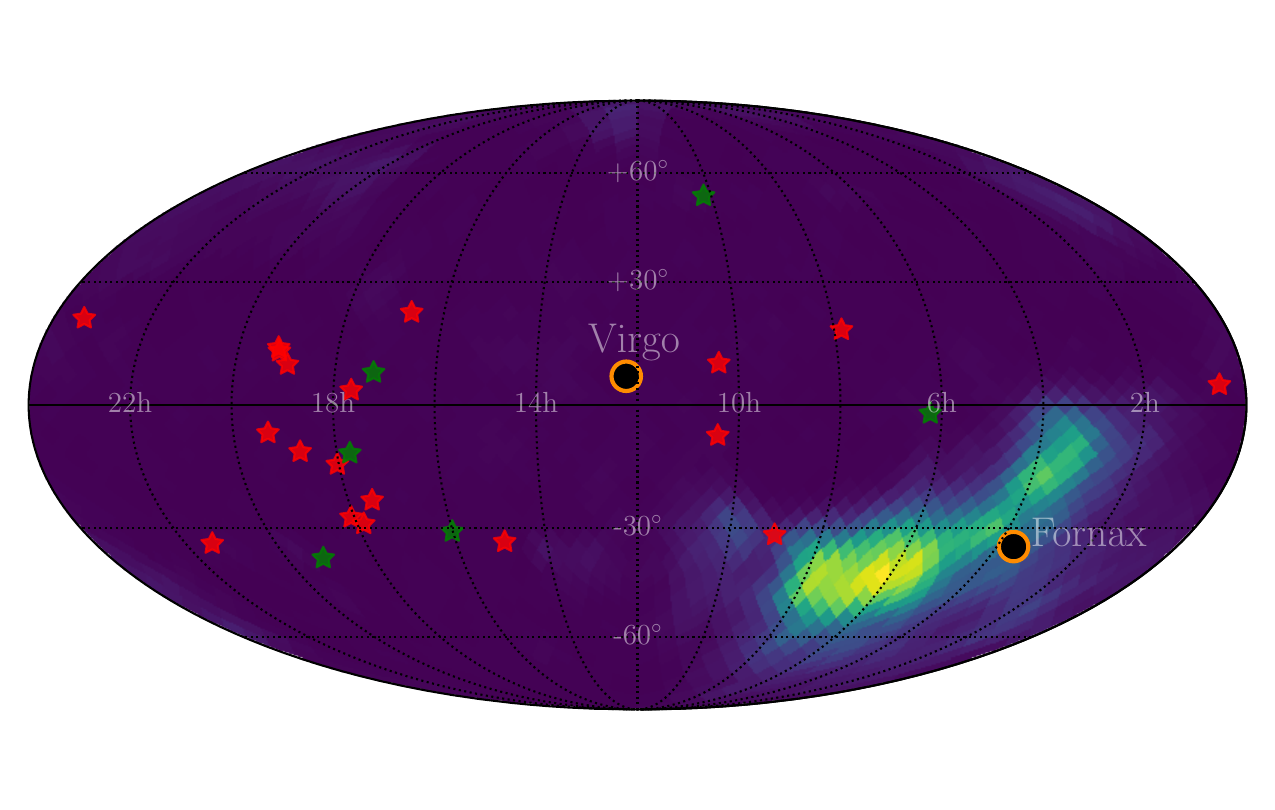}
\caption{Posterior distribution of the sky localization obtained by searching for a CGW in the second data release of the EPTA \texttt{DR2new}. The posteriors are obtained using a CGW model with Earth and pulsar term, the inclusion of the custom pulsar noise (PSRN) and a common uncorrelated red noise (CURN) represented by 9 frequency bins. For reference, we show the position of the analyzed pulsars and the Virgo and Fornax clusters.}
\label{fig:curn_cgw_sky_posterior}
\end{figure*}

The simplest considered data model includes only the custom pulsar noise (PSRN), therefore, no CRN is included. The PSRN model is used as a null hypothesis, and the alternative is given by the pulsar noise plus CGW (PSRN+CGW) considering Earth and pulsar terms. The Bayes factor for the model comparison PSRN+CGW vs PSRN is 4000.

Next, we include a CRN to the custom noise PSRN: a CURN or a GWB correlated according to the HD pattern. These become the new null hypotheses (CURN+PSRN) and (GWB+PSRN). We also consider two descriptions for the CRN: one using the three lowest Fourier harmonics (3 bins) and one using the nine lowest Fourier harmonics (9 bins) as in~\citep{wm3}. For reference, the maximum resolvable frequency for three and nine bins are $\log_{10}(3/(10.3 yr))\approx -8.03$ and $\log_{10}(9/(10.3 yr))\approx -7.56$, respectively. In practice, we should let the data choose the number of frequency bins similarly to what was done in \cite{wm2}. However, for practical reasons, we present the results only for 3 and 9 bins. The maximum resolvable frequencies smaller than the one over-a-year frequency dictate the choice of nine frequency bins. Since the power-law model adopted to describe the background has two parameters, the minimum number of frequencies we can fit is three. Therefore, we also present the model selection results for a choice of three bins. Note that the free spectrum analysis presented in \citep{wm3} shows that the first, fourth, and ninth bins are constrained with a tail extending to low amplitudes. Only the second bin is constrained with zero support at low powers. Since the CGW candidate is located close to the second Fourier bin, showing the results for three bins can help to single out the red noise components of the spectrum that might be potentially affected by the other high-frequency noises. The Bayes factors of PSRN+CURN+CGW vs PSRN+CURN are 4 and 12, for 9 and 3 bins, respectively.
The choice of the number of bins affects the spectral properties of the CRN and, consequently, also the Bayes factors. In fact, the slope of the CURN model becomes steeper when using 3 bins, allowing a possible CGW to emerge from the noise. When including the HD correlations, the Bayes factors of PSRN+GWB+CGW vs PSRN+GWB drop to 0.7 and 1, for 9 and 3 bins, respectively. 

As already was pointed out in \citep{wm3}, the HD component of the noise absorbs most of CGW signal and this can be clearly seen in the drop of the Bayes factor and in the posterior distributions shown in \autoref{fig:hd_curn_posterior}. When the CRN is a CURN, the CGW model absorbs the power of the background around $\log_{10}f_{gw}\in[-8.5,-8.2]$ and yields an amplitude $\log_{10}A$ posterior distribution with tails extending up to the lowest end of the prior range (see correlation in \autoref{fig:hd_curn_posterior} for parameters $\log_{10}f_{gw},\log_{10}A$).

For the model CURN+PSRN+CGW with 9 bins, the log-frequency $f_{gw}$ is measured to be $4.61^{+1.11} _{-2.98}$ nHz and the log-amplitude $\log_{10} h$ is measured to be $-14.0^{+0.5} _{-2.6}$ (median and symmetric 90\% credible interval). The chirp-mass posterior is uninformative and the sky localization posterior is shown in \autoref{fig:curn_cgw_sky_posterior} where we also show the Virgo and Fornax clusters which are $ 16.5$~Mpc and $ 19.3$~Mpc from the Earth \cite{Jordan:2007aw}, respectively. If we use the median values of the amplitude and frequency to estimate the luminosity distance, we obtain $d_L\approx 16.6 \, ({\mathcal{M}}/{10^9 {\rm M}_{\odot}} )^{5/3} \,{\rm Mpc},$ using \autoref{eq:h_mc_f}. 

We compute the sky marginalized 95\% upper limit on strain amplitude, $h_{95}(f_{gw})$, across the studied frequency range for the model PSRN+CURN+CGW with 9 bins. For this analysis, we used a uniform prior on $h$ in the range $[10^{-18}, 10^{-11}]$ instead of the uniform prior on $\log_{10}h$ used for the search (see \cite{arzoumanian2023nanograv, Falxa_2023}). The strain upper limit was converted into a horizon distance, $D_H$, (i.e., the distance up to which SMBHB systems should produce detectable CGW signals) using \autoref{eq:h_mc_f}:
\begin{equation}
    D_H = 2 \frac{\mathcal{M}^{5/3}}{h_{95}}(\pi f_{gw})^{2/3}.
\label{eq:horizon_distance}
\end{equation}

\begin{figure}
\centering
\includegraphics[width=0.98\linewidth]{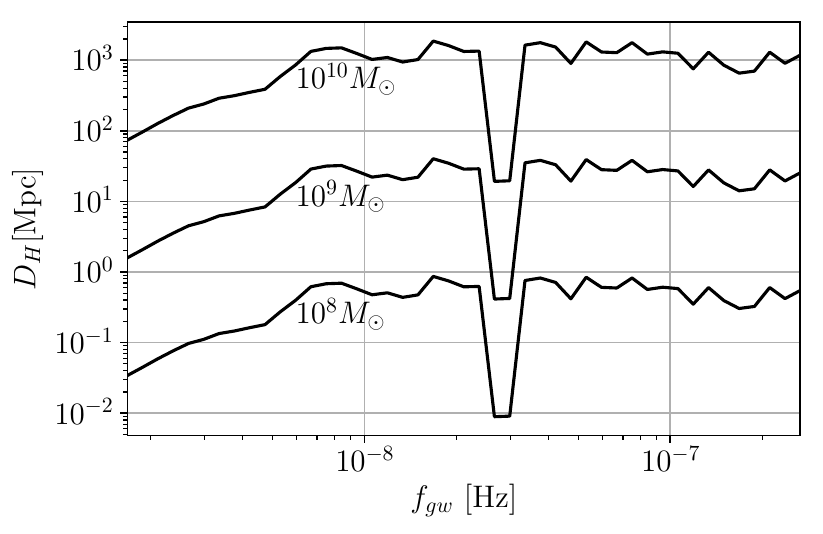}
\caption{Horizon luminosity distance, $D_H$, obtained from the sky averaged 95\% upper limit on strain amplitude $h$ using the PSRN+CURN+CGW (Earth term + pulsar term) model. The horizon distance is calculated with \autoref{eq:horizon_distance} for three chirp masses: $10^8 M_\odot$, $10^9 M_\odot$ and $10^{10} M_\odot$.
For reference, the Virgo and Fornax clusters are at $ 16.5$~Mpc and $ 19.3$~Mpc from the Earth, respectively.
}
\label{fig:horizon_distance}
\end{figure}

We plot $D_H$ as a function of $f_{gw}$ in \autoref{fig:horizon_distance} for three values of chirp mass $\mathcal{M} = [10^8 M_\odot, 10^9 M_\odot, 10^{10} M_\odot ]$. The highest $D_H$ is recovered around 20 nHz. The closest galaxy-cluster candidates (Fornax and Virgo) that could host a SMBHB lie at distances larger than 10 Mpc, meaning that we need binary systems with chirp masses larger than $10^9 M_\odot$ in order for them to be detectable. One of the known galaxies hosting a supermassive black hole in the Fornax cluster is the radio galaxy Fornax A (NGC 1316). The radio observations suggest the presence of a powerful AGN, and thus a central supermassive black hole, whose estimated mass is $M=1.5\times 10^{8} \, M_\odot$ \citep{Nowak:2008wu}. The mass and distance of such a black hole make it outside of the horizon distance of our PTA experiment.

\begin{figure*}
\centering
\includegraphics[width=0.8\linewidth]{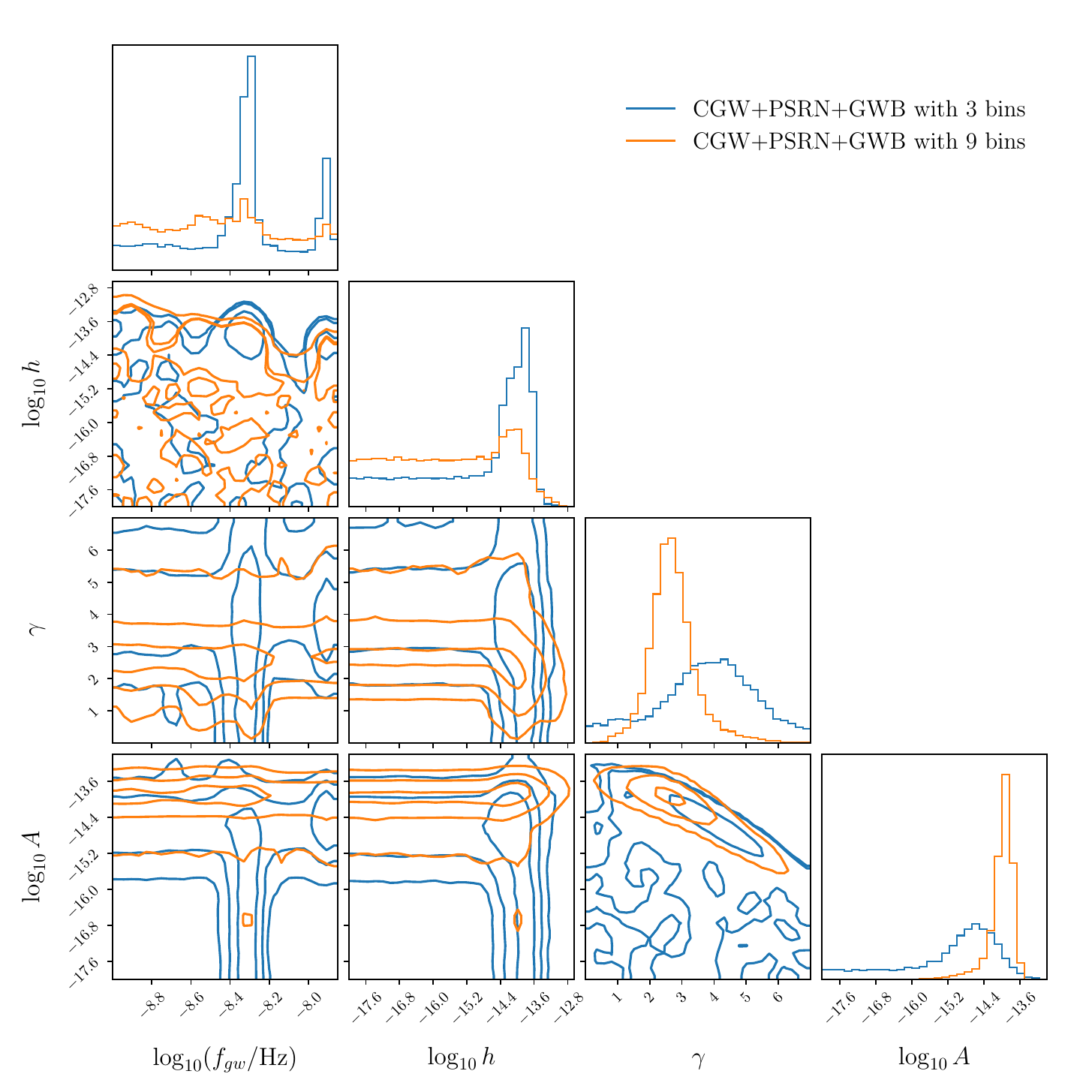}
\caption{Posterior distributions of the CGW search in the second data release of the EPTA \texttt{DR2new}. The posteriors are obtained using a CGW model with Earth and pulsar term, the custom PSRN model and an HD correlated background (GWB) represented by 9 and 3 frequency bins. We show the posterior distribution for the gravitational wave frequency and amplitude $f_{gw}, h$ of the CGW, and the common noise spectral index and amplitude $\gamma$ and $A$. The contours indicate the 1,2,3-$\sigma$ Gaussian contours.}
\label{fig:cgw_gwb_bin_comparison}
\end{figure*}

For the model GWB+PSRN+CGW with 9 bins, we cannot constrain the CGW parameters, so we set a 95\% upper limit on $\log_{10} h_{95\%} = -13.75$. We constrain the spectral properties of the GWB background to be $2.66 _{-1.02} ^{+1.43}$ and $-13.95_{-0.62} ^{+0.25}$, respectively for $\gamma$ and $\log_{10}A)$ (median and symmetric 90\% credible interval).  In \autoref{fig:cgw_gwb_bin_comparison} we show the posteriors for the same model GWB+PSRN+CGW, but this time with 9 and 3 bins. 
The background's spectral properties ($\gamma$ and $A$) are affected by the number of bins. The median log-amplitude decreases from -13.95 to -14.34 and the median slope from 2.66 to 3.832, following the typical $\gamma-\log_{10}A$ correlation. The steeper slope allows the CGW to emerge from the noise and its posteriors show two clear peaks, one at 4.64 nHz and one at 12.6 nHz. Since the GWB model with 3 bins extends in frequency up to $9.33$ nHz, the appearance of the second peak at 12.6 nHz might be due to unmodelled noise.

\begin{figure}
\centering
\includegraphics[width=0.98\linewidth]{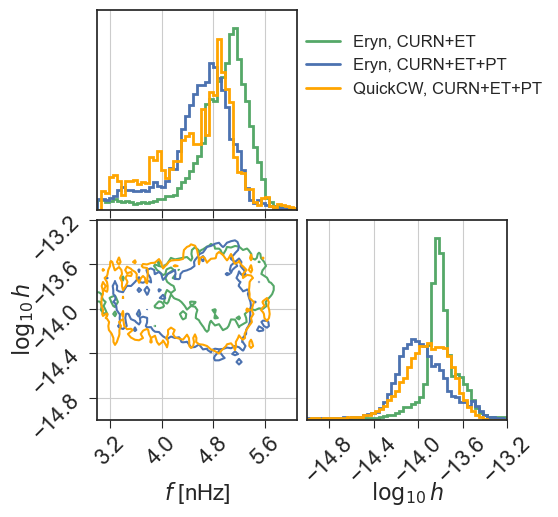}
\caption{Inference of the amplitude $h$ and frequency $f_{gw}$ of CGW using the PSRN+CURN+CGW model. The results obtained with \texttt{Eryn} are shown as green and blue histograms, for a Earth term only, and full CGW model, respectively. The blue contours are to be compared with the orange posterior obtained with \texttt{QuickCW}.
The shown contours are the 90\% normal credible regions.
}
\label{fig:ET_PT_sampl}
\end{figure}

The inclusion of the HD correlation absorbs the presence of the CGW signal. We still need to check whether the peak frequency at 4.6 nHz for a CURN background is unaffected by the sampler choice, the pulsar term or the inclusion of eccentricity. To do this, we consider the model CURN+PSRN+CGW. The narrow green posterior contours in \autoref{fig:ET_PT_sampl} correspond to using \texttt{Eryn} to sample the model with the Earth-term only and the broad blue posterior is inferred with the model including the pulsar term. We have overplotted similar results (including the pulsar term) obtained with the \texttt{QuickCW} sampler in orange. The results are in broad agreement independently of the sampler choice and the exclusion of the pulsar term.
To check if the circular CGW model is appropriate, we carried out a separate analysis including orbital eccentricity in the model following \cite{Taylor:2015kpa} but using only the Earth term in the analysis. This analysis inferred a low eccentricity ($e<0.2$) for the CGW candidate, indicating that the analysis performed with the circular binary model is adequate.

\subsection{Optimal Statistics}

The Bayesian analysis presented in the previous subsection indicates that the results about the nature of the observed signal are inconclusive. This subsection starts a long investigation process attempting to answer the question: ``what is it that we see?'' and provides a transition to the next section which describes analysis of the simulated data. 

\begin{figure*}
\centering
\includegraphics[width=.75\textwidth]{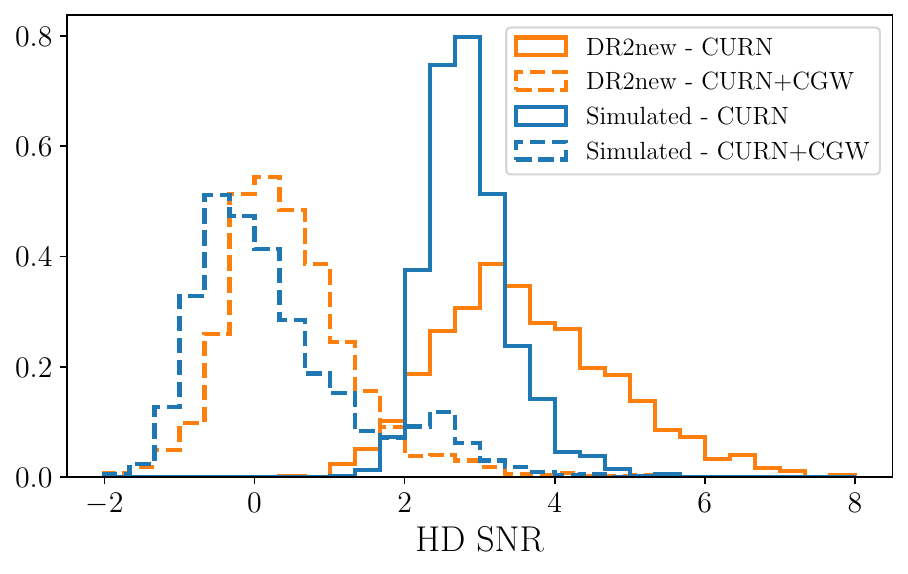}
\caption{Distributions of the signal-to-noise-ratio (SNR) for the HD correlations for a common red process with $\gamma = 13/3$, with (dashed) or without (solid) adding CGW to the data model. The orange lines correspond to the results obtained on \texttt{DR2new} and blue are on the simulated data with only CGW (no CRN).
}
\label{fig:OS_SNR}
\end{figure*}

We compute the signal-to-noise ratio (SNR) of the CRN using the optimal statistic approach \citep{PhysRevD.91.044048, 2018PhRvD..98d4003V} implemented within \texttt{enterprise\_extensions} software package \citep{enterprise_extensions}. We estimate the SNR assuming quadrupolar correlation (HD). Following the procedure outlined in \citep{wm3, 2018PhRvD..98d4003V}, we vary the pulsar noise parameters (using the custom noise models from \citep{wm2})
and get as a result a distribution of SNR. The pulsar noise parameters are drawn from the posterior samples of a Bayesian analysis, which includes, in addition to single pulsar noises, a CURN, using a 9 Fourier bin basis. The solid orange line in \autoref{fig:OS_SNR} reproduces the findings reported in \citep{wm3}.

Next we include a CGW in the model: we use all the posterior samples obtained in the Bayesian analysis which preserve the correlation between the noise parameters and the CGW (instead of only the noise parameters), and re-evaluate the optimal statistic. As a result, for each posterior sample, the SNR is computed, given the sample pulsar noise parameters, after subtracting the deterministic CGW residuals from the pulsar residuals. The CGW residuals are computed using the sample CGW parameter values. The resulting SNR distribution of HD correlations is given in  \autoref{fig:OS_SNR} as dashed orange line. This result implies that the data (minus CGW) does not show any sign of quadrupolar (GWB) correlation, in other words, a CGW alone can explain the HD feature observed in the data.

We corroborate our results obtained on the EPTA \texttt{DR2new} by repeating the same analysis on a simulated dataset.  
 We produce a fake PTA based on the real (EPTA \texttt{DR2new}) pulsars and the noise estimation in which we inject only one CGW and no GWB (see Section~\ref{sec:simulations} for a detailed description of the simulation).  The blue solid line in \autoref{fig:OS_SNR} 
 indeed resembles the result obtained on the \texttt{DR2new} (orange line). As we will discuss in detail in the next section, a single CGW signal could be interpreted as a GWB (see \cite{Allen_2023}). As expected, the subtraction of the CGW from the timing residuals removes the quadrupolar correlation (blue dashed line) and reproduces the previously obtained results on the EPTA data.

\section{Simulation}
\label{sec:simulations}

We perform a simulation campaign to try to reproduce the features observed in the analysis of \texttt{DR2new}. We generate a fake array with the same time of arrivals (TOA)s and pulsar positions as in the real dataset. We inject noises using the maximum a posteriori of the noise parameter posterior obtained in \cite{wm2}. We use a Gaussian process to simulate the noise components and consider different realizations in order to reproduce the observed results (see Appendix~\ref{sec:appendix_simulation}). Using the simulated array as a basis, we propose two cases to study:

\begin{itemize}
    \item \texttt{PSRN+CGW}: A simulated analogue of \texttt{DR2new} with only one circular CGW injected at 4.8 nHz with sky location at (3h38, -35$^\circ$27) as if it was in the Fornax cluster with a chirp mass of $10^{9.2} M_\odot$ and amplitude $h=10^{-13.6}$, without any CRN.
    \item \texttt{PSRN+GWB}: A simulated analogue of \texttt{DR2new} with a gaussian and isotropic GWB as CRN with a powerlaw spectrum corresponding to $A = 10^{-14.5}$ and spectral index $\gamma=13/3$ (without any CGW).
\end{itemize}

Each simulation is analysed with the custom PSRN model and either with CGW (using the Earth term only) or GWB with a powerlaw spectrum.

\begin{figure}
\centering
\includegraphics[width=0.98\linewidth]{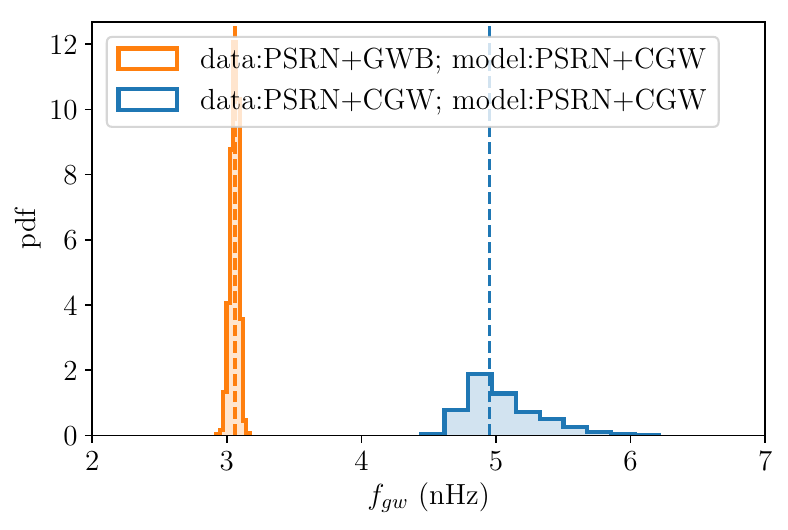}
\caption{Posterior distribution of the gravitational wave frequency $f_{gw}$ of a CGW fitted to two simulated PTAs: one with a single injected CGW (\texttt{PSRN+CGW} blue histogram), and one with an injected GWB (\texttt{PSRN+GWB} orange histogram). The posterior distribution is obtained with a MCMC analysis for a PSRN+CGW (Earth term) model. The dashed lines are the medians of the distributions.}
\label{fig:sim_freq}
\end{figure}

\begin{figure}
\centering
\includegraphics[width=0.98\linewidth]{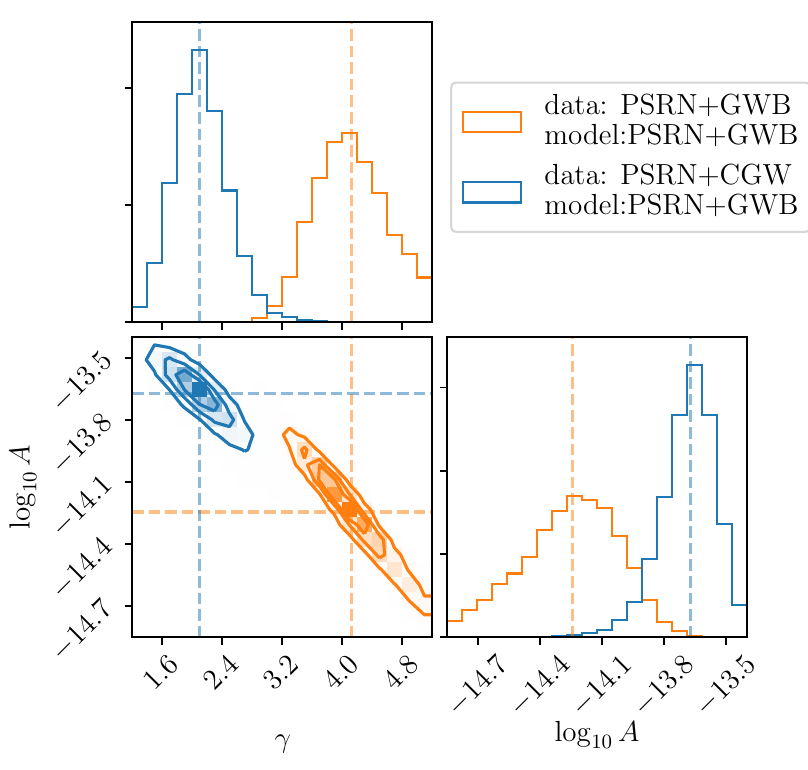}
\caption{Posterior distribution of amplitude $\log_{10}A$ and spectral index $\gamma$ of the GWB for two simulated PTAs : 1 realistic PTA with only 1 CGW injected (\texttt{PSRN+CGW}) and 1 realistic PTA with a GWB injected (\texttt{PSRN+GWB}). The posterior distribution is obtained with a MCMC analysis for a PSRN+GWB model. The dashed lines are the medians of the distributions.}
\label{fig:sim_gwb}
\end{figure}

We have considered the \texttt{PSRN+GWB} simulated data and analysed it with a single CGW source (no GWB). In \autoref{fig:sim_freq} we show that we can recover the CGW even if we have injected an isotropic GWB. We have repeated the analysis on 10 simulated datasets, in all cases the recovered ``CGW'' was centered at the lowest Fourier bin ($1/T_{obs} \sim 3$ nHz) and located in the close vicinity of pulsars, in many cases around J1713+0747. Recovering the lowest frequency bin is compatible with the injected powerlaw spectrum for which it is the loudest bin. We present the posterior frequency of one case as a orange histogram in \autoref{fig:sim_freq}. We have also analysed this data using GWB model and the inferred posterior is given in orange in \autoref{fig:sim_gwb}. 

Next, we consider the \texttt{PSRN+CGW} simulated data and analyse it with the model of isotropic GWB with a power-law spectrum. As expected, this model gives a constrained posterior. Indeed, the excess of power at low frequency gives support to a power-law spectrum, though not the best description because we try to fit a single frequency process with a continuous spectrum. The HD correlations could be reproduced as shown in \autoref{fig:OS_SNR}. A single source can generate the HD pattern by averaging over the pulsar pairs (see \cite{Allen_2023, Cornish_2013} for details). In the present case, the SNR is dominated by a few pulsars so the averaging is strongly biased and we would not necessarily expect a strong HD component. The analysis of this \texttt{PSRN+CGW} dataset with a CGW model is shown in \autoref{fig:sim_freq} as a blue histogram. Analysis of the same data with GWB is given in \autoref{fig:sim_gwb} as a blue posterior. 
One can see that it has a higher amplitude and is shallower,  similar to what was presented in \cite{wm3}.

The analysis of the simulated data demonstrates how a CGW model can mimic a GWB and vice-versa. The point source also produces HD correlations, as previously shown in \citep{Cornish_2013, B_csy_2022, Allen_2023}. Moreover, the anisotropic configuration of the current PTA (pulsars not uniformly distributed in the sky and having very different noise properties) produces an uneven response across the sky and the studied frequency range (see \autoref{subsec:fstat_results}). The interactions between our configuration of pulsars and the signal models is a challenging aspect of PTAs that has been investigated in previous studies \citep{2022MNRAS.509.5538C, 2023MNRAS.518.1802S}. The inclusion of pulsars from the southern hemisphere (PPTA) would provide a better coverage of the sky allowing a better localization of single sources, which is crucial to differentiate between a GWB or a CGW.

\section{Summary}
\label{sec:conclusions}

This paper presents an analysis of the EPTA \texttt{DR2new} dataset searching for continuous GW signals from super-massive black hole binaries in quasi-circular orbits. We perform a frequentist (based on $\mathcal{F}_e$-statistic) and Bayesian (using Bayes factor) analysis of the data, and, in both cases, find a significant CGW candidate at 4-5.6 nHz. The frequentist analysis gives a p-value of ($5\times 10^{-4} $ -- $6\times 10^{-3}$), equivalent to a 2.5-3$\sigma$ significance level, depending on the evaluation procedure and whether or not a CURN is included in the noise model. Within the Bayesian analysis of the CGW candidate, we computed the Bayes factor to assess whether the data prefers the inclusion of a CGW source. We find weak evidence (Bayes factors $\sim 4-12$) for the inclusion of a CGW on top of a CURN process. If the common red noise is assumed to have the HD correlation typical of a GWB, the Bayes factors for the inclusion of a CGW drop to $\sim 0.7-1$. The data is equally well described by a model including both a GWB and CGW and a model including GWB only. Even though the CGW model depends on 58 parameters and therefore comes with a large dimensionality penalty, the Bayes factor is close to unity. More data will allow to make these hypothesis test more conclusive. 

In an attempt to understand if the observed signal is due to a GWB or a CGW, we perform a simulation campaign. We simulate data based on the noise parameters inferred in \cite{wm2} and inject a GW signal. Our main finding is that simulated data with only an isotropic GWB injected can be fitted with a CGW model, and vice versa; a GWB model can explain simulated data containing only a single injected CGW. Therefore, we cannot conclusively distinguish between the presence of a single continuous gravitational wave or a gravitational wave background. In \cite{interpretation_paper}, considering models that produce a GW signal consistent with the one present in \texttt{DR2new}, the probability of detecting a single source with SNR larger than 3 is estimated to be 50\%.

We hope that an analysis of the combined IPTA data (Data Release 3) will help to confirm the presence or not of a CGW signal and shed light on its nature.

\begin{acknowledgements}
The European Pulsar Timing Array (EPTA) is a collaboration between
European and partner institutes, namely ASTRON (NL), INAF/Osservatorio
di Cagliari (IT), Max-Planck-Institut f\"{u}r Radioastronomie (GER),
Nan\c{c}ay/Paris Observatory (FRA), the University of Manchester (UK),
the University of Birmingham (UK), the University of East Anglia (UK),
the University of Bielefeld (GER), the University of Paris (FRA), the
University of Milan-Bicocca (IT), the Foundation for Research and 
Technology, Hellas (GR), and Peking University (CHN), with the
aim to provide high-precision pulsar timing to work towards the direct
detection of low-frequency gravitational waves. An Advanced Grant of
the European Research Council allowed to implement the Large European Array
for Pulsars (LEAP) under Grant Agreement Number 227947 (PI M. Kramer). 
The EPTA is part of the
International Pulsar Timing Array (IPTA); we thank our
IPTA colleagues for their support and help with this paper and the external Detection Committee members for their work on the Detection Checklist.

Part of this work is based on observations with the 100-m telescope of
the Max-Planck-Institut f\"{u}r Radioastronomie (MPIfR) at Effelsberg
in Germany. Pulsar research at the Jodrell Bank Centre for
Astrophysics and the observations using the Lovell Telescope are
supported by a Consolidated Grant (ST/T000414/1) from the UK's Science
and Technology Facilities Council (STFC). ICN is also supported by the
STFC doctoral training grant ST/T506291/1. The Nan{\c c}ay radio
Observatory is operated by the Paris Observatory, associated with the
French Centre National de la Recherche Scientifique (CNRS), and
partially supported by the Region Centre in France. We acknowledge
financial support from ``Programme National de Cosmologie and
Galaxies'' (PNCG), and ``Programme National Hautes Energies'' (PNHE)
funded by CNRS/INSU-IN2P3-INP, CEA and CNES, France. We acknowledge
financial support from Agence Nationale de la Recherche
(ANR-18-CE31-0015), France. The Westerbork Synthesis Radio Telescope
is operated by the Netherlands Institute for Radio Astronomy (ASTRON)
with support from the Netherlands Foundation for Scientific Research
(NWO). The Sardinia Radio Telescope (SRT) is funded by the Department
of University and Research (MIUR), the Italian Space Agency (ASI), and
the Autonomous Region of Sardinia (RAS) and is operated as a National
Facility by the National Institute for Astrophysics (INAF).

The work is supported by the National SKA programme of China
(2020SKA0120100), Max-Planck Partner Group, NSFC 11690024, CAS
Cultivation Project for FAST Scientific. This work is also supported
as part of the ``LEGACY'' MPG-CAS collaboration on low-frequency
gravitational wave astronomy. JA acknowledges support from the
European Commission (Grant Agreement number: 101094354). JA and SCha 
were partially supported by the Stavros
Niarchos Foundation (SNF) and the Hellenic Foundation for Research and
Innovation (H.F.R.I.) under the 2nd Call of the ``Science and Society --
Action Always strive for excellence -- Theodoros Papazoglou''
(Project Number: 01431). AC acknowledges support from the Paris
\^{I}le-de-France Region. AC, AF, ASe, ASa, EB, DI, GMS, MBo acknowledge
financial support provided under the European Union's H2020 ERC
Consolidator Grant ``Binary Massive Black Hole Astrophysics'' (B
Massive, Grant Agreement: 818691). GD, KLi, RK and MK acknowledge support
from European Research Council (ERC) Synergy Grant ``BlackHoleCam'', 
Grant Agreement Number 610058. SB, HQ acknowledge funding from the French National Research Agency (grant ANR-21-CE31-0026, project MBH\_waves). This work is supported by the ERC 
Advanced Grant ``LEAP'', Grant Agreement Number 227947 (PI M. Kramer). 
KLi acknowledges support from the Shanghai Astronomical Observatory, Chinese Academy of Sciences, 80 Nandan Road, Shanghai 200030, China.
AV and PRB are supported by the UK's Science
and Technology Facilities Council (STFC; grant ST/W000946/1). AV also acknowledges
the support of the Royal Society and Wolfson Foundation. JPWV acknowledges
support by the Deutsche Forschungsgemeinschaft (DFG) through thew
Heisenberg programme (Project No. 433075039) and by the NSF through
AccelNet award \#2114721. NKP is funded by the Deutsche
Forschungsgemeinschaft (DFG, German Research Foundation) --
Projektnummer PO 2758/1--1, through the Walter--Benjamin
programme. ASa thanks the Alexander von Humboldt foundation in
Germany for a Humboldt fellowship for postdoctoral researchers. APo, DP and MBu acknowledge support from the INAF Large Grant 2022 “GCjewels” (P.I. Andrea Possenti) approved with the Presidential Decree 30/2022
(Italy). RNC acknowledges financial support from the Special Account
for Research Funds of the Hellenic Open University (ELKE-HOU) under
the research programme ``GRAVPUL'' (K.E.-80383/grant agreement 319/10-10-2022: PI N. A. B. Gizani). EvdW, CGB and GHJ acknowledge support from the Dutch National Science
Agenda, NWA Startimpuls – 400.17.608.
BG is supported by the Italian Ministry of Education, University and 
Research within the PRIN 2017 Research Program Framework, n. 2017SYRTCN. LS acknowledges the use of the HPC system Cobra at the Max Planck Computing and Data Facility.

\ifnum\wm>1 The Indian Pulsar Timing Array (InPTA) is an Indo-Japanese
collaboration that routinely employs TIFR's upgraded Giant Metrewave
Radio Telescope for monitoring a set of IPTA pulsars.  BCJ, YG, YM,
SD, AG and PR acknowledge the support of the Department of Atomic
Energy, Government of India, under Project Identification \# RTI 4002.
BCJ, YG and YM acknowledge support of the Department of Atomic Energy,
Government of India, under project No. 12-R\&D-TFR-5.02-0700 while SD,
AG and PR acknowledge support of the Department of Atomic Energy,
Government of India, under project no. 12-R\&D-TFR-5.02-0200.  KT is
partially supported by JSPS KAKENHI Grant Numbers 20H00180, 21H01130,
and 21H04467, Bilateral Joint Research Projects of JSPS, and the ISM
Cooperative Research Program (2021-ISMCRP-2017). AS is supported by
the NANOGrav NSF Physics Frontiers Center (awards \#1430284 and
2020265).  AKP is supported by CSIR fellowship Grant number
09/0079(15784)/2022-EMR-I.  SH is supported by JSPS KAKENHI Grant
Number 20J20509.  KN is supported by the Birla Institute of Technology
\& Science Institute fellowship.  AmS is supported by CSIR fellowship
Grant number 09/1001(12656)/2021-EMR-I and T-641 (DST-ICPS).  TK is
partially supported by the JSPS Overseas Challenge Program for Young
Researchers.  We acknowledge the National Supercomputing Mission (NSM)
for providing computing resources of ‘PARAM Ganga’ at the Indian
Institute of Technology Roorkee as well as `PARAM Seva' at IIT
Hyderabad, which is implemented by C-DAC and supported by the Ministry
of Electronics and Information Technology (MeitY) and Department of
Science and Technology (DST), Government of India. DD acknowledges the 
support from the Department of Atomic Energy, Government of India 
through Apex Project - Advance Research and Education in Mathematical 
Sciences at IMSc. \fi

The work presented here is a culmination of many years of data
analysis as well as software and instrument development. In particular,
we thank Drs. N.~D'Amico, P.~C.~C.~Freire, R.~van Haasteren, 
C.~Jordan, K.~Lazaridis, P.~Lazarus, L.~Lentati, O.~L\"{o}hmer and 
R.~Smits for their past contributions. We also
thank Dr. N. Wex for supporting the calculations of the
galactic acceleration as well as the related discussions.
The EPTA is also grateful to staff at its observatories and telescopes who have made the continued observations possible.

\linebreak\linebreak\textit{Author contributions.}
The EPTA is a multi-decade effort and all authors have
contributed through conceptualisation, funding acquisition,
data-curation, methodology, software and hardware
 developments as well as (aspects of) the continued running of
the observational campaigns, which includes writing and
proofreading observing proposals, evaluating observations
and observing systems, mentoring students, developing
science cases. All authors also helped in (aspects of)
verification of the data, analysis and results as well as
in finalising the paper draft. Specific contributions from individual 
EPTA members are listed in the CRediT\footnote{\url{https://credit.niso.org/}} format below.

We especially thank K. Grunthal for having found a bug in the implementation of the F-statistics.

InPTA members contributed in uGMRT observations and data reduction to
create the InPTA data set which is employed while assembling the
\texttt{DR2full+} and \texttt{DR2new+} data sets. 

\ifnum\wm=1

JJan, KLi, GMS equally share the correspondence of the paper.

\linebreak\linebreak\textit{CRediT statement:}\newline
Conceptualisation: APa, APo, AV, BWS, CGB, CT, GHJ, GMS, GT, IC, JA, JJan, JPWV, JW, JWM, KJL, KLi, MK.\\
Methodology: APa, AV, DJC, GMS, IC, JA, JJan, JPWV, JWM, KJL, KLi, LG, MK.\\
Software: AC, AJ, APa, CGB, DJC, GMS, IC, JA, JJan, JJaw, JPWV, KJL, KLi, LG, MJK, RK.\\
Validation: AC, APa, CGB, CT, GMS, GT, IC, JA, JJan, JPWV, JWM, KLi, LG.\\
Formal Analysis: APa, CGB, DJC, DP, EvdW, GHJ, GMS, JA, JJan, JPWV, JWM, KLi.\\
Investigation: APa, APo, BWS, CGB, DJC, DP, GMS, GT, IC, JA, JJan, JPWV, JWM, KLi, LG, MBM, MBu, MJK, RK.\\
Resources: APa, APe, APo, BWS, GHJ, GMS, GT, HH, IC, JA, JJan, JPWV, JWM, KJL, KLi, LG, MJK, MK, RK.\\
Data Curation: AC, AJ, APa, BWS, CGB, DJC, DP, EG, EvdW, GHJ, GMS, GT, HH, IC, JA, JJan, JPWV, JWM, KLi, LG, MBM, MBu, MJK, MK, NKP, RK, SChe, YJG.\\
Writing – Original Draft: APa, GMS, JA, JJan, KLi, LG.\\
Writing – Review \& Editing: AC, AF, APa, APo, DJC, EB, EFK, GHJ, GMS, GT, JA, JJan, JPWV, JWM, KLi, MK, SChe, VVK.\\
Visualisation: APa, GMS, JA, JJan, KLi.\\
Supervision: APo, ASe, AV, BWS, CGB, DJC, EFK, GHJ, GMS, GT, IC, JA, JPWV, KJL, KLi, LG, MJK, MK, VVK.\\
Project Administration: APo, ASe, AV, BWS, CGB, CT, GHJ, GMS, GT, IC, JJan, JPWV, JWM, KLi, LG, MK.\\
Funding Acquisition: APe, APo, ASe, BWS, GHJ, GT, IC, JA, JJan, LG, MJK, MK.\\

\fi

\ifnum\wm=2
InPTA members contributed to the discussions that probed the impact of 
including InPTA data on single pulsar noise analysis. Furthermore, they 
provided quantitative comparisons of various noise models, wrote a brief 
description of the underlying \texttt{Tensiometer} package, and helped 
with the related interpretations.

SB, MF, LS equally share the correspondence of the paper. 

\linebreak\linebreak\textit{CRediT statement:}\newline
Conceptualisation: AC, APa, APo, AV, BWS, CT, GMS, GT, JPWV, JWM, KJL, KLi, MJK, MK.\\
Methodology: AC, APa, AV, DJC, GMS, IC, JWM, KJL, KLi, LG, MJK, MK, SB, SChe, VVK.\\
Software: AC, AJ, APa, APe, GD, GMS, KJL, KLi, MJK, RK, SChe, VVK.\\
Validation: AC, APa, BG, GMS, IC, JPWV, JWM, KLi, LG, MJK.\\
Formal Analysis: AC, APa, BG, EvdW, GHJ, GMS, JWM, KLi, MJK.\\
Investigation: AC, APa, APo, BWS, CGB, DJC, DP, GMS, IC, JPWV, JWM, KLi, LG, MBM, MBu, MJK, RK, VVK.\\
Resources: AC, APa, APe, APo, BWS, GHJ, GMS, GT, IC, JPWV, JWM, KJL, KLi, LG, MJK, MK, RK.\\
Data Curation: AC, AJ, APa, BWS, CGB, DJC, DP, EvdW, GHJ, GMS, JA, JWM, KLi, MBM, MJK, MK, NKP, RK, SChe.\\
Writing – Original Draft: AC, APa, GMS, MJK.\\
Writing – Review \& Editing: AC, AF, APa, APo, BG, EB, EFK, GMS, GT, JA, JPWV, JWM, KLi, MJK, MK, SChe, VVK.\\
Visualisation: AC, APa, GMS, KLi, MJK.\\
Supervision: AC, APo, ASe, AV, BWS, CGB, DJC, EFK, GHJ, GT, JPWV, KJL, LG, MJK, MK, VVK.\\
Project Administration: AC, APo, ASe, AV, BWS, CGB, CT, GHJ, GMS, GT, JPWV, JWM, LG, MJK, MK.\\
Funding Acquisition: APe, APo, ASe, BWS, GHJ, GT, IC, LG, MJK, MK.\\
\fi

\ifnum\wm=4

Additionally, InPTA members contributed to GWB search efforts with 
\texttt{DR2full+} and \texttt{DR2new+} data sets and their interpretations. 
Further, they provided quantitative comparisons of GWB posteriors that 
arise from these data sets and multiple pipelines.

For this work specifically, SB, MF, LS equally share the 
correspondence of the paper. 

\linebreak\linebreak\textit{CRediT statement:}\newline
Conceptualization	Conceptualization: AShe, av, GMS, HQL, LS, LSe, MF, MK.\\
Methodology	Methodology: HQL, IF, JG, JWMo, LS, LSe, MF, MK.\\
Software	Software: AC, GMS, HQL, JWMo, LS, LSe, MF, MJK, SC.\\
Validation	Validation: DI, HQL, IF, JWMo, LS, LSe, MF.\\
Formal Analysis	Formal Analysis: GMS, HQL, IF, JG, LS, LSe, MF.\\
Investigation	Investigation: DP, HQL, I, LS, LSe, MBM, MF, SAS.\\
Resources	Resources: AC, AP, GJu, GT, I, JPWVi, KL, LS, LSe, MF, MJK, MK.\\
Data Curation	Data Curation: AC, DP, GMS, HH, I, KL, LS, LSe, MBa, MBM, MF, MJK, SAS, SC.\\
Writing – Original Draft	Writing – Original Draft: GMS, HQL, LS, LSe, MF.\\
Writing – Review \& Editing	Writing – Review \& Editing: AC, AFo, av, GMS, HQL, JG, KL, LS, LSe, MF, SC.\\
Visualization	Visualization: HQL, KL, LS, LSe, MF.\\
Supervision	Supervision: AP, AShe, av, GT, JG, JPWVi, LS, LSe, MF.\\
Project Administration	Project Administration: av, GT, JPWVi, LS, LSe, MF, MK.\\
Funding Acquisition	Funding Acquisition: AP, AShe, av, GT, I, MJK, MK.\\
\fi
\end{acknowledgements}

%
%

\bibpunct{(}{)}{;}{a}{}{,}
\def\aap{A\&A}                
\def\aapr{A\&A~Rev.}          
\def\aaps{A\&AS}              
\def\aj{AJ}                   
\def\ajph{Australian J.~Phys.}
\def\alet{Astro.~Lett.}       
\def\ao{Applied Optics}       
\def\apj{ApJ}                 
\def\apjl{ApJ}                
\def\apjs{ApJS}              
\def\apss{Ap\&SS}             
\def\araa{ARA\&A}             
\def\asr{Av.~Space Res.}     
\def\azh{AZh}                 
\def\baas{BAAS}               
\def\cpc{Comput.~Phys.~Commun.} 
\def\gca{Geochim.~Cosmochim.~Acta} 
\def\iaucirc{IAU Circ.}       
\def\ibvs{IBVS}               
\def\icarus{Icarus}           
\def\jcomph{J.~Comput.~Phys.} 
\def\jcp{J.~Chem.~Phys.}      
\def\jgr{J.~Geophys.~R.}      
\def\jrasc{JRASC}             
\def\met{Meteoritics}         
\def\mmras{MmRAS}             
\def\mnras{MNRAS}             
\def\mps{Meteoritics and Planetary Science} 
\def\nast{New Astron.}        
\def\nat{Nature}              
\def\pasj{PASJ}               
\def\pasp{PASP}               
\def\phr{Phys.~Rev.}          
\def\pdra{Phys.~Rev.~A}       
\def\prb{Phys.~Rev.~B}       
\def\prc{Phys.~Rev.~C}       
\def\prd{Phys.~Rev.~D}       
\def\phrep{Phys.~Rep.}        
\def\phss{Phys.~Stat.~Sol.}        %
\def\procspie{Proc.~SPIE}     
\def\planss{Planet.~Space Sci.}  
\def\qjras{QJRAS}             
\def\rpph{Rep.~Prog.~Phys.}   
\def\rgsp{Rev.~Geophys.~Space Phys.~} 
\def\sal{Soviet Astron.~Lett.}
\def\sci{Science}             
\def\solph{Sol.~Phys.}        
\def\ssr{Space Sci.~Rev.}     
\def\zap{Z.~Astrophys.}       
\def\jasa{J.~Amer.~Stat.~Assoc.} 

\bibliographystyle{aa}
\bibliography{psrrefs,references}

\begin{appendix}

\section{Simulation}
\label{sec:appendix_simulation}

The simulations were performed using the \texttt{fakepta} package (\url{https://github.com/mfalxa/fakepta}). The injected noises are assumed to be stationary gaussian processes with a power spectral density $S_n(f, {\alpha})$ function of frequency $f$ and hyper-parameters ${\alpha}$ according to \cite{wm2}. In time domain, such noises $n({t})$ can be decomposed on a Fourier basis of size $N$:

\begin{equation}
    n({t}) = \sum_{k=1} ^N X_k \sin(2\pi k {t} / T_{obs}) + Y_k \cos(2\pi k {t} / T_{obs}) = \textbf{F} {a},
\end{equation}
where we rewrote the sum as a matrix-vector multiplication
with ${a}$ a vector of $(X_k, Y_k)$ and $\textbf{F}$ the design matrix of size $N_{\rm TOAs} \times N$ containing the sine and cosine terms $\sin(2\pi k {t} / T_{obs}),\cos(2\pi k {t} / T_{obs})$ per each TOA.

The covariance matrix $\textbf{C}_n ({t}, {t'})$ of the gaussian process $n(t)$ is given by the expectation value $\langle \rangle$ of the noise at two times:
\begin{equation}
    \textbf{C}_n ({t}, {t'}) = \langle n({t}), n({t'}) \rangle = \textbf{F}^\intercal \langle {a}^2 \rangle \textbf{F} = \textbf{F}^\intercal {\Phi} \textbf{F}
\end{equation}
with ${\Phi} = \textrm{diag} \big\{ S_n(k/T_{obs})/T_{obs} \big\}$.

Considering that the stochastic process $n({t})$ follows a multivariate gaussian distributions $\mathcal{N}({0}, \textbf{C}_n)$, a random realisation of $n({t})$ corresponds to a random draw from the distribution $\mathcal{N}({0}, \textbf{C}_n)$. The noises are injected by summing the obtained $n({t})$ to the simulated pulsar residuals for the maximum a posteriori set of noise hyper-parameters ${\alpha}$ (entering in the $S_n$) inferred from the real data.

\end{appendix}

\end{document}